\newif\ifconfver
\confverfalse      %declaring conference version false
\ifconfver
     \documentclass[10pt,twocolumn,twoside]{IEEEtran}
\else
    \documentclass[12pt,draftcls,onecolumn]{IEEEtran}
\fi

\usepackage{calc,amsfonts,amssymb,amsmath,bm,url,color,theorem,cite}
\usepackage{graphicx,subfigure}
\usepackage{epsfig,multirow, epstopdf}
\usepackage{algorithmic,algorithm}

\newlength{\twidth}
\ifconfver
\else
\fi

    \makeatletter
    \def\multilimits@{\bgroup
  \Let@
  \restore@math@cr
  \default@tag
 \baselineskip\fontdimen10 \scriptfont\tw@
 \advance\baselineskip\fontdimen12 \scriptfont\tw@
 \lineskip\thr@@\fontdimen8 \scriptfont\thr@@
 \lineskiplimit\lineskip
 \vbox\bgroup\ialign\bgroup\hfil$\m@th\scriptstyle{##}$\hfil\crcr}
    \def\Sb{_\multilimits@}
    \def\endSb{\crcr\egroup\egroup\egroup}
\makeatother

\newtheorem{Lemma}{Lemma}
\newtheorem{Prop}{Proposition}
\newtheorem{Theorem}{Theorem}

\theorembodyfont{\rmfamily}

\definecolor{orange}{RGB}{255,107,0}
\definecolor{purple}{rgb}{0.627,0.125,0.941}

{\begin{list}{}%
    {\setlength{\rightmargin}{0pc}%
    \setlength{\leftmargin}{1pc} }} %
{\end{list}}

{\begin{list}{}%
    {\setlength{\rightmargin}{1pc}%
    \setlength{\labelsep}{0pc}
    \setlength{\leftmargin}{1pc} }} %
{\end{list}}

{\begin{list}{}%
    {\setlength{\rightmargin}{0pc}%
    \setlength{\leftmargin}{0pc} }} %
{\end{list}}

{\begin{list}{}{
    \settowidth{\labelwidth}{\mbox{\textnormal{#1}}}%
    \setlength{\leftmargin}{\labelwidth+\labelsep}}}%
{\end{list}}

\begin{document}

\bibliographystyle{IEEEtran}

\title{Semidefinite Relaxation and Approximation Analysis of a Beamformed Alamouti Scheme for Relay Beamforming Networks}
%\title{Strategies for Exploring Two Degrees-of-freedom \\in Relay Beamforming Networks and the Performance Analysis}

\ifconfver \else {\linespread{1.1} \rm \fi

\author{
\ifconfver
Sissi Xiaoxiao Wu, Anthony Man-Cho So, Jiaxian Pan and Wing-Kin Ma
    \thanks{This work was supported in part by the Hong Kong Research Grant Council (RGC) General Research Fund (GRF) Project CUHK 416012, and in part by The Chinese University of Hong Kong Direct Grant No. 2050506.}
    \thanks{Sissi Xiaoxiao Wu is the corresponding author. She is now with the Signal, Information, Networks and Energy Laboratory, Arizona State University, Tempe, United States. E-mail: xxwu.eesissi@gmail.com.
		Anthony Man-Cho So is with the
    Department of Systems Engineering and Engineering Management, CUHK, Hong Kong.  E-mail: manchoso@ee.cuhk.edu.hk. Jiaxian Pan and Wing-Kin Ma are with the Department of Electronic Engineering, The Chinese University of Hong Kong (CUHK), Hong Kong.
    E-mail: jxpan@ee.cuhk.edu.hk, wkma@ieee.org.}
\else
Sissi Xiaoxiao Wu, Anthony Man-Cho So, Jiaxian Pan and Wing-Kin Ma
    %\thanks{Copyright (c) 2012 IEEE. Personal use of this material is permitted.
    %However, permission to use this material for any other purposes must be obtained from the IEEE by sending a request to pubs-permissions@ieee.org.}
    \thanks{This work was supported in part by the Hong Kong Research Grant Council (RGC) General Research Fund (GRF) Project CUHK 416012, and in part by The Chinese University of Hong Kong Direct Grant No. 2050506.}
    \thanks{Sissi Xiaoxiao Wu is the corresponding author. She is now with the Signal, Information, Networks and Energy Laboratory, Arizona State University, Tempe, United States. E-mail: xxwu.eesissi@gmail.com.
		Anthony Man-Cho So is with the
    Department of Systems Engineering and Engineering Management, CUHK, Hong Kong.  E-mail: manchoso@ee.cuhk.edu.hk. Jiaxian Pan and Wing-Kin Ma are with the Department of Electronic Engineering, The Chinese University of Hong Kong (CUHK), Hong Kong.
    E-mail: jxpan@ee.cuhk.edu.hk, wkma@ieee.org.}
\fi
}

\maketitle

\ifconfver \else
%\begin{center} \vspace*{-3\baselineskip}
%Final Version,
%%\today
%March 2015
%\\[\baselineskip]
%\end{center}
\fi

\begin{abstract}
In this paper, we study the amplify-and-forward (AF) schemes in two-hop one-way relay networks. In particular, we consider the multigroup multicast transmission between long-distance users. Given that perfect channel state information is perceived, our goal is to design the AF process so that the max-min-fair (MMF) signal-to-interference-plus-noise ratio (SINR) is optimized subject to generalized power constraints.  We propose a rank-two beamformed Alamouti (BFA) AF scheme and formulate the corresponding AF design problem as a \emph{two-variable} fractional quadratically-constrained quadratic program (QCQP), which is further tackled by the semidefinite relaxation (SDR) technique. We analyze the approximation quality of two-variable fractional SDRs under the Gaussian randomization algorithm. These results are fundamentally new and reveal that the proposed BFA AF scheme can outperform the traditional BF AF scheme, especially when there are many users in the system or many generalized power constraints in the problem formulation.  From a practical perspective, the BFA AF scheme offers two degrees of freedom (DoFs) in beamformer design, as opposed to the one DoF offered by the BF AF scheme, to improve the receivers' SINR. In the latter part of this paper, we demonstrate how this extra DoF leads to provable performance gains by considering two special cases of multicasting, where the AF process is shown to employ a special structure. The numerical simulations further validate that the proposed BFA AF scheme outperforms the BF AF scheme and works well for large-scale relay systems.

\ifconfver  \else
\\[.5\baselineskip]
\noindent
{\bfseries EDICS}: %MSP-CODR (MIMO precoder/decoder design),
SPC-PERF (Performance analysis and bounds)
MSP-APPL (Applications of MIMO communications and signal processing),
SAM-BEAM (Applications of sensor and array multichannel processing)
%MSP-STCD  (MIMO space-time coding and capacity),
%MSP-CAPC  (MIMO capacity and performance)
\fi
\end{abstract}

%\begin{keywords}
%secrecy capacity, transmit beamforming, Charnes-Cooper
%transformation, convex optimization, semi-definite program
%(SDP).\end{keywords}
%
%%\ifconfver \else
%%   \vspace*{\baselineskip}
%%   {\bfseries EDICS}:
%%   {MSP-DECD} (MIMO space-time coding and decoding algorithms),
%%   SPC-DETC (Detection, estimation, and demodulation)
%%\fi

\ifconfver \else } \fi

\ifconfver
\else
\newpage
\fi

%==================================================
%-----------------------------------------------------------------------------

\section{Introduction}\label{sec:intro}
\begin{figure}[h]
\centering
  \includegraphics[width=0.5\textwidth]{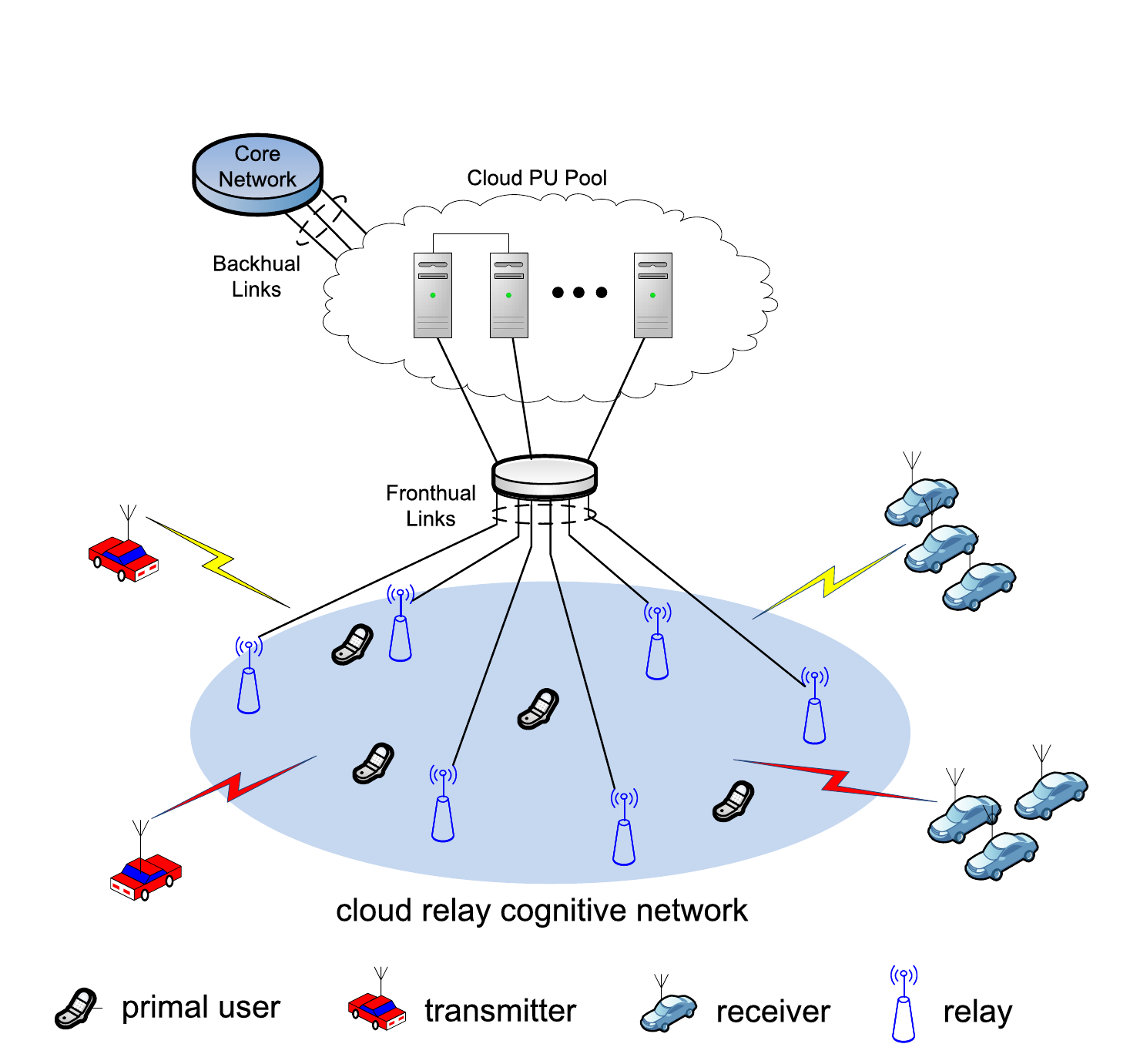}
\caption{An example of the cloud relay network.}
\label{fig:cloud_relay}
\end{figure}

The information delivery between multiple wireless devices has shown an increasing importance in up-to-date military networks, relay networks, and 5G networks \cite{ ITAWIFI, tehrani2014device, cloudrelay0, li2014social}. The state-of-the-art technique in this context is to use small smart access points (APs), e.g., mobile phones, wireless relays, Wi-Fi APs, to assist information delivery between far-apart transceiver pairs. 
Nowadays,  a new trend is to connect the smart APs by fibers, microwave or millimeter wave to build up a cloud processing center for facilitating reliable communications. A typical example is the cloud radio access network (C-RAN) \cite{wpcran, shi2013group, cranfronthaul}, which is recently proposed as a promising network architecture to offer a 1000x increase in capacity to support broadband applications. The key enabling technologies in C-RANs are the cloud processors pool and fronthaul-backhaul links. They coordinate all the base-stations in all cells as a cloud base-station and serve the users in a jointly optimized way.  In view of this, it is promising to extend the ``cloud relays" by considering the communications between devices as information delivery in a relay network. We treat the intra-network interference as noise and try to handle the inter-network interference by designing the AF process in the cloud center.
%We call all the intermediate APs as relays, no matter they are smart phones, Wi-Fi APs or any other kinds of relay APs. Note that in cell networks, the base-station or other users can cause interference to the D2D communication. 
%We may avoid the interference by a proper scheduling or just simply treat the interference as noise.} 
This gives rise to the so-called \emph{cloud relay network} (C-RN) \cite{cloudrelay0, wucrn_r2, WuLiMaSo2015SPAWC, wu2015relaysbf}; see a system model example in Figure \ref{fig:cloud_relay}. 

In this work, we are interested in a typical two-hop one-way relay network. In particular, we consider the case that all the nodes, i.e., the transmitter, the receiver, and relays, are equipped with a single antenna. This assumption is reasonable, as nodes in a D2D communication network are usually limited by power and apparatus. In our setting, the transmitter and receiver are far-apart, and thus they rely on the relays to amplify and forward (AF) the information,\footnote{The relays can also decode-and-forward (DF) the received signals, but this is beyond the scope of this paper.} since
the direct link between them are negligible. We assume that the relays are distributively located, and more importantly, they coordinate to form a C-RN. In practice, the capacity of the fronthaul-backhaul links in C-RNs is an important issue. If the link capacity is unlimited, both the channel state information (CSI) and received signals can be shared within the cloud, and thus the system becomes an MIMO relay network. On the other hand, if the link capacity is limited in such a way that only CSIs are shared and the received signals are isolated among different relays, then the system is reduced to a distributed relay network. We remark that there are also other kinds of relay networks corresponding to different link capacities, but in this paper we focus on the two above. 

In this paper, we are interested in relay AF design for multigroup multicast transmission. In the literature, there are different problem formulations; see, e.g.,~\cite{fazeli2009multiple,Jnl:Relay_Chalise_09,chalise2007mimo, 
%distributed_relay_Ding_08, bornhorst2012distributed, havary2008distributed, distributed_relay_Goma07, Cooperative_relay_Poor11, 
%chae2008mimo,Non_Regenerative_MIMO,Heath_MIMOrelay_08,
%Jnl:Relay_Chalise_09,Linear_MMSE,Precoder_Design,jimenez2012non,khandaker2012joint,MCY12,taowang12,chamaeh13,choi2014transceiver,
%dual_hop_MISO_relay2014,
distributed_relay_Ding_08,havary2008distributed,jimenez2012non,distributed_relay_Goma07, khandaker2012joint, KR14}. Herein, we focus on the max-min-fair (MMF) formulation, in which the worst user's signal-to-interference-plus-noise ratio (SINR) is to be maximized under generalized power constraints, such as total power constraint, per-relay power constraints, interference temperature constraints, or energy harvesting constraints. This makes our design approach applicable to many scenarios.
A classic approach for AF relays is to adopt the beamformed (BF) AF scheme \cite{fazeli2009multiple, Jnl:Relay_Chalise_09
}, which gives rise to an NP-hard \emph{one-variable fractional} quadratically-constrained quadratic program (QCQP). An efficient way to tackle QCQP is to apply the semidefinite relaxation (SDR) technique~\cite{Jnl:MagzineMaLUO}. That is, we first rewrite the fractional QCQP to a rank-one constrained fractional semidefinite program (SDP). Then, by dropping the non-convex rank-one constraint, we solve the SDP in polynomial time. It is well known that the SDR is tight if the corresponding SDP has rank-one solutions. Otherwise, a Gaussian randomization algorithm is applied to convert the SDP optimal solution into a feasible solution to the fractional QCQP~\cite{chang2008approximation,jimulti13}. We shall call this solution the \emph{SDR solution} in the sequel. A fundamental issue here is to quantify the quality of the SDR solution.  In our previous work \cite{WuLiMaSo2015SPAWC, wu2015relaysbf}, we show that the SDP always has a rank-one solution when $M+J \le 4$, and that the SINR associated with the SDR solution is at least $\Omega(\frac{1}{{M\log J}})$ times that associated with the optimal solution to the fractional QCQP when $M+J>4$, where $M$ is the number of users (receivers) in the network and $J$ is the number of generalized power constraints. This result provides an SDR approximation bound for the one-variable fractional QCQP when multiple constraints are present. However, from a practical perspective, it actually implies that the SINR associated with the SDR-based BF AF scheme may experience a performance loss on the order of $1/({M\log J})$ in large-scale systems.
%, where there are many users served in the network or many generalized power constraints in the problem formulation. 

In order to improve the relay beamforming performance, we propose to adopt the Alamouti space-time code in the AF structure.  This leads to the \emph{BF Alamouti (BFA) AF} scheme, in which two beamformers are used to process two data symbols jointly.  Compared to the BF AF scheme, which uses only one beamformer to process a single data symbol, the BFA AF scheme has one extra degree of freedom (DoF) in the beamformer design.  As such, it is expected that the latter will yield better system performance.  In fact, the extra DoF available to the BFA AF scheme is also manifested in its corresponding design problem, in that it can be shown to admit a \emph{two-variable} fractional QCQP formulation. Our analytical results show that the optimal value of the corresponding two-variable SDP is always no worse than that of the one-variable SDP.  In addition, the SDR is tight when $M + J \le 5$, and when $M+J>5$, the approximation quality of the SDR-based BFA AF scheme is on the order of $1/(\sqrt{M}\log J)$.  Clearly, both the tightness and approximation quality results are better than their BF AF counterparts.

The idea of using the Alamouti code in the context of single-group multicast beamforming has been introduced in our previous work~\cite{MainPaper}, where the design problem is a standard QCQP. The subsequent conference paper \cite{jimulti13} proposed the BFA scheme for multigroup multicasting without relays. This paper significantly improves \cite{jimulti13} by generalizing the aforementioned BFA scheme to relay networks with generalized power constraints and providing an analysis on its approximation performance. It is worth mentioning that the authors in \cite{schad2012convex} also introduced a rank-two BF AF scheme for relay networks. However, the scenario they focused on is single-group multicasting in a distributed relay network with one power constraint, while our work considers multigroup multicasting for both the distributed relay and MIMO relay networks with generalized power constraints. We remark that the generalization from single-group multicasting to multigroup multicasting, and from distributed relays to MIMO relays, are non-trivial. More importantly, our work establishes for the first time the SDR approximation quality of a fairly general class of two-variable fractional QCQPs.  It should also be noted that the problem considered in this paper, namely beamformer design for multi-user to multi-user multigroup multicasting in relay networks, has not been well addressed in the literature.  Indeed, existing works on relay transceiver design mainly focus on the point-to-point~\cite{distributed_relay_Ding_08, havary2008distributed},
%, Cooperative_relay_Poor11, MCY12,dual_hop_MISO_relay2014,Non_Regenerative_MIMO,Linear_MMSE,Precoder_Design},
single-user to multi-user~\cite{jimenez2012non}, multi-user to single-user~\cite{distributed_relay_Goma07, khandaker2012joint}, and multi-user to multi-user unicast~\cite{fazeli2009multiple}
%, choi2014transceiver},
%Heath_MIMOrelay_08,chalise2007mimo,Jnl:Relay_Chalise_09} 
and multicast scenarios~\cite{KR14}. The work in \cite{schad2015rank} studies the rank-two relay multicasting with a direct link, while our work is targeted at far-apart devices without direct links and our scenario setting is the more generalized multigroup multicasting. 
The work in~\cite{bornhorst2012distributed} studies the beamformer design in a multigroup multicast relay network.  However, it considers BF AF schemes for single-antenna relays, whereas our focus is on rank-two BFA AF scheme under a cloud relay setting. We remark that some efficient heuristics have recently been proposed to find a high-quality solution to one-variable fractional QCQPs; see, e.g., \cite{tran2014conic,christopoulos2015multicast,mehanna2015feasible,Gopalakrishnan15}. We numerically compare their results with the proposed BF Alamouti AF scheme in the the companion technical report~\cite{CompTechRepr2}, wherein we could see that even though those heuristics can help find a better QCQP solution, the proposed BFA AF still owns a significantly better performance.

The organization of this paper is as follows. In Section~II, we introduce the system model and review the SDR-based BF AF scheme for both the MIMO relay network and distributed relay network in the presence of primal users. We show how the design problem for the BF AF scheme can be tackled by the SDR technique and introduce the existing approximation bounds on the performance of the SDR solution. In Section~III, we introduce the BFA AF scheme and show how the corresponding design problem gives rise to a new two-variable fractional QCQP formulation. As one of our main results, we establish an approximation bound for the SDR of the two-variable fractional QCQP, thereby providing a performance guarantee for the proposed BFA AF scheme. In Section IV, special cases of multicasting relay networks are studied, from which we see how the two DoFs offered by the BFA AF scheme can improve the system performance. Lastly, simulation results are provided in Section~\ref{sec:sim},
and the paper is concluded in Section~\ref{sec:conclusions}.

%\noindent
%{\it Notations:}
Our notation is standard:
$\mathbb{C}^N$ is the set of all complex $N$-dimensional vectors;
%$\mathbb{H}^N$ is the set of all $N \times N$ complex Hermitian matrices;
$\mathbb{H}_+^{N}$  is the set of all $N \times N$ positive semidefinite matrices;
%${\bf x} \ge {\bf 0}$ means that ${\bf x}$ is elementwise non-negative;
%${\bf X} \succeq {\bf 0}$ means that ${\bf X}$ is positive semidefinite;
$\| \cdot \|$ is the vector Euclidean norm;
${\bm A} \bullet {\bm B}$ stands for the inner product between matrix ${\bm A}$ and ${\bm B}$; 
${\bm A} \otimes  {\bm B}$ stands for the Kronecker product between matrix ${\bm A}$ and ${\bm B}$;
${\bm A} \odot  {\bm B}$ stands for the element-wise product between matrix ${\bm A}$ and ${\bm B}$;
${\rm vec}({\bm A})$ is the vectorization operator for matrix ${\bm A}$; 
${\rm Diag}({\bf x})$ is a diagonal matrix parametrized with the elements of ${\bf x}$;
%${\rm rank}({\bm X})$, ${\rm \lambda}_{\rm max}({\bm X})$, ${\rm \lambda}_{\rm min}^+({\bm X})$ and ${\rm \lambda}_{1}({\bm X})$
%stand for the rank, the largest eigenvalue, the smallest non-zero eigenvalue of ${\bm X}$ and the smallest eigenvalue of ${\bm X}$, resp.;
${\rm rank}({\bm X})$ and ${\rm \lambda}_{\min}({\bm X})$ stand for the rank and the smallest eigenvalue, resp.;
%$\mathcal{R}({\bm X})$ denotes the range space of ${\bm X}$;
%${\bf 0}$ and  ${\bf 1}$ are the all-zero and all-one vectors, resp.;
${\bf e}_i$ is a unit vector with the nonzero element at the $i$th entry;
%${\bf 1}$ is an all-one vector;
${\bm I}_r$ denotes the $r$-by-$r$ identity matrix;
%{\em\color{red} \{ the above all being matrix things; below are statistical \} }
$\mathbb{E}[ \cdot ]$ is statistical expectation;
$\mathcal{CN}({\bf 0},{\bm W})$ is used to denote the circularly symmetric complex Gaussian distribution  with mean vector ${\bf 0}$ and covariance matrix ${\bm W}$. 
%and ${\bm X} \sim {\bm Y}$ means that the random variables ${\bm X}$ and ${\bm Y}$ have the same distribution.
%{\em\color{red} \{ add whenever we introduce new notations \} }

%==================================================
%-----------------------------------------------------------------------------
\section{System Model and the Beamformed Amplify-and-Forward Schemes}
In this section, we describe the system model for two-hop one-way relay networks. In particular, we consider the \emph{multigroup multicast} transmission by a network of single-antenna AF relays. We assume that there are $L$ relays in the network; $G$ single-antenna transmitters (sources) send $G$ independent common information to $G$ groups of single-antenna users (receivers, destinations); users in the same group require the same information while users in different groups require different information. In total there are $\sum_{j=1}^Gm_j=M$ users in the network, where $m_j$ is the number of users in group $j,~\forall j=1,...,G$. In our target setting, the transmitters and receivers are far-apart such that the direct link between them can be ignored. Thus, relays play an important role during the information delivery by amplifying and forwarding (AF) received signals from sources to destinations.  By introducing the cloud relay network (C-RN) concept \cite{wucrn_r2}, we consider that the relays are distributively located but connected by a cloud PUs pool (i.e., the computation center) via fronthaul and backhaul links, which are typically fibers or microwave connected to fibers. We assume that all nodes in the network are well synchronized, channels from transmitters to relays and relays to users are frequency flat, quasi-static, and the channels are perfectly perceived by transmitters sending reference signals and then passed to the PUs pool. The PUs pool can serve as a genie in the network to design the AF process.

\subsection{The Beamformed AF Scheme}
Let us first introduce the traditional AF scheme, i.e., the BF AF scheme, which is widely used in existing relay networks, e.g.,~\cite{Jnl:Relay_Chalise_09, fazeli2009multiple}. We begin with the information delivery process in a one-way relay network, which is proceed by two hops:

\noindent
1) Source-to-relay hop: \emph{the transmitters send information to relays.} The receive model at the relay is written as follows
\begin{equation}\label{rt}
{\bf r}(t) = \sum_{j=1}^G{\bm f}_js_j(t) + {\bf n}(t),
\end{equation}
where ${\bf r}(t) = [r^1(t),..., r^\ell(t),...,r^L(t)]^T$ with 
\begin{equation}\label{rt_v}
{r}^\ell(t) = \sum_{j=1}^G{f}_j^\ell s_j(t) + {n}^\ell(t), \quad  \ell = 1,...,L
\end{equation}
being the received signal at relay $\ell$; $s_j(t)$ is the common information specific for group $j$ with $\mathbb{E}[|s_j(t)|^2]=P_j$, where $P_j$ is the transmit power at transmitter $j$; ${\bm f_j}= [f_j^1,...,f_j^\ell,...,f_j^L]^T$ is the channel from transmitter $j$ to all the relays;  ${\bf n}(t) = [{n}^1(t),...,{n}^\ell(t),...,{n}^L(t)]^T$ with ${n}^\ell(t)$ being the Gaussian noise at relay $\ell$ which has mean zero and variance $\sigma_\ell^2$. Throughout this paper, we assume that $\sigma_\ell^2>0, \forall \ell$.

\noindent
2) Relay-to-destination hop: \emph{relays process the received signals and then forward them to receivers.} In particular, the transmit structure at the relay side can be given by
\begin{equation} \label{eq:xt_v}
{\bf x}(t) = {\bm V} {\bf r}(t).
\end{equation}
The weighting matrix ${\bm V}$ represents how the relays interact with each other. Specifically,
we may consider all the single-antenna relays are connected by a cloud, within which the CSIs are fully shared. 
Then, if the received signals are shared within the cloud, ${\bm V}$ could be any arbitrary matrix to be optimized, and we are dealing with an MIMO relay network~\cite{Jnl:Relay_Chalise_09}; if the received signals are not shared, ${\bm V}$ is a diagonal matrix, and we are dealing with a distributed relay network~\cite{fazeli2009multiple}. 

Let us first consider the MIMO relay network. The received signal for user-$(k, i)$, i.e., user $i$ in group $k$, is given by
\begin{align}\label{yt}
y_{k, i}(t) &= {\bm g}_{k, i}^H{\bf x}(t) + {v_{k, i}}(t) \\\notag
%= & {\bm g}_{k, i}^H{\bm V}\left(\sum_{j=1}^G{\bm f}_js_j(t)\right) + {\bm g}_{k, i}^H{\bm V}{\bf n}(t) + {v_{k, i}}(t) \\\notag
&= \underbrace{{\bm g}_{k, i}^H{\bm V}{\bm f}_ks_k(t)}_{{\rm desired~signal}} + \\\notag
&\underbrace{{\bm g}_{k, i}^H{\bm V}\left(\sum_{m \neq k}{\bm f}_ms_m(t)\right)+ {\bm g}_{k, i}^H{\bm V}{\bf n}(t) + {\mu_{k, i}}(t)}_{{\rm interference~and~noise}},
\end{align}
where ${\bm g}_{k, i} = [g_{k, i}^1,...,g_{k, i}^\ell,...,g_{k, i}^L]^T $ is the channel from relays to user-$(k, i)$ and ${\mu_{k, i}}(t)$ is the Gaussian noise at user-$(k, i)$ with zero mean and variance $\sigma_{{k, i}}^2$. In this paper, we assume that $\sigma_{k, i}^2>0, \forall k, i$.
Accordingly, the receive SINR at user-$({k,i})$ can be expressed as
\begin{align}\label{gamma}
\frac{P_k\left|{\bm g}_{k, i}^H{\bm V}{\bm f}_k\right|^2}{\displaystyle\sum_{m \neq k}P_m\left|{\bm g}_{k, i}^H{\bm V}{\bm f}_m\right|^2+{\bm g}_{k,i}^H{\bm V}{\bm \Sigma}{\bm V}^H {\bm g}_{k,i}+ \sigma_{k,i}^2},
\end{align}
where ${\bm \Sigma} = {\rm Diag}(\sigma_1^2,...,\sigma_\ell^2,...,\sigma_L^2) \succ {\bf 0}$.

\subsection{Generalized Power Constraints}
Our goal is to optimize the AF process for the relay networks. Specifically, to design the AF weighting matrix ${\bm V}$, three design constraints are under our considerations:
\begin{itemize}
\item[1).] \emph{Total Power Constraints. }  A natural consideration is that the total transmit power at relays is below a given threshold. In this way, we have the total power constraint given by
\begin{align}\notag
\mathbb{E}[\|{\bf x}(t)\|^2]=&{\rm Tr} \left({\bm V}\left(\sum_{j=1}^{G} P_j{\bm f}_j{\bm f}_j^H+ {\bm \Sigma}\right){\bm V}^H \right) \\\label{d_v1}
=&{\bm w}^H {\bm D}_0{\bm w}\le \bar P_0,
\end{align}
where \eqref{d_v1} is obtained by denoting ${\bm w} = {\rm vec}({\bm V})$ and using the identity
$$
{\rm Tr}\left({\bm A}^H{\bm B}{\bm C}{\bm D}\right)= {\rm vec}({\bm A})^H \left( {\bm D}^T \otimes {\bm B} \right) {\rm vec}({\bm C}),
$$
which is valid for arbitrary complex matrices ${\bm A}, {\bm B}, {\bm C}, {\bm D}$ of appropriate dimensions. Also, 
$\bar P_0$ is the total transmit power budget for all the relays and ${\bm D}_0$ is defined in \eqref{d_v0}
in Table \ref{tab:BFAF_summary}.
%which is obtained by using the following equality
%\begin{equation} \label{vec}
%{\rm Tr}\left({\bm A}^H{\bm B}{\bm C}{\bm D}\right)= {\rm vec}({\bm A})^H({\bm D}^T \otimes {\bm B}){\rm vec}({\bm C}).
%\end{equation}
\item[2).] \emph{Per-relay Power Constraints.} Another common design constraint is subject to the power allowed at each relay, which can be written as
\begin{align*} \label{eq_per-power}
{\bf e}_{\ell}^H{\bm V}\left(\sum_{j=1}^{G} P_j{\bm f}_j{\bm f}_j^H+ {\bm \Sigma}\right){\bm V}^H {\bf e}_{\ell}  \le  \bar P_\ell, \quad \ell=1,...,L,
\end{align*}
where $\bar P_\ell$ is the maximum transmit power allowed at relay $\ell$. 
Similar to \eqref{d_v1}, we can further rewrite
it to
\begin{equation} \label{d_v2}
\quad {\bm w}^H {\bm D}_\ell{\bm w}\le \bar P_\ell, \quad \ell=1,...,L,
\end{equation}
where
${\bm D}_\ell$ is defined in \eqref{d_v}
in Table \ref{tab:BFAF_summary}.

\item[3).] \emph{Interference Temperature Constraints. } We may also consider a popular design constraint in the cognitive radio (CR) network is to control the interference temperature at primal users~\cite{huang2012robust, phan2009spectrum, zhang2011optimal}. We consider a CR relay network with $U$ primal users where the interference is simply caused by the AF relays (not the transmitters). Let ${\bm h}_{u} = [h_{u}^1,...,h_{u}^\ell,...,h_{u}^L]^T$ be the channel from relays to primal user $u$, $\forall u=1,..., U$.  Then the interference temperature power constraints for primal users are given by
\begin{equation*}\label{cont_cr}
\quad {\bm w}^H {\bm G}_u{\bm w}\le  \eta_u, \quad u=1,...,U,
\end{equation*}
where $\eta_u$ is the maximum interference allowed at primal user $u$, ${\bm G}_u$ is defined in \eqref{G_u}
in Table \ref{tab:BFAF_summary} with $\sigma_{u}^2>0$ being the noise power at primal user $u$.

 %\item[4).] \emph{Energy Harvesting Constraints.} Recently, there is flourishing interest in using RF wave to transfer power for those users in the system that aim for receiving energy rather than information. We usually call this kind of users as \emph{energy receivers} (ERs). Assuming that ${\bm r}_{s} = [r_{s}^1,...,r_{s}^\ell,...,r_{s}^L]^T$ is the channel from relays to ER-$s$, we could impose a group of harvested energy constraints at the ER side \cite{Varshney,Grover,Zhang}:
%\begin{align}\label{cont_er}
%\mu_s{\bm w}^H {\bm R}_s {\bm w} \le \nu_s, \quad j=s,...,S,
%\end{align}
%where ${\bm R}_s$ is defined in \eqref{R_s}
%in Table \ref{tab:BFAF_summary},
%$0< \mu_j <1$ denotes the energy harvesting efficiency at the ER $j$~\cite{le2008efficient,agbinya2012wireless} and $\nu_j$ is required power threshold at the ER-$j$.
\end{itemize}
%In the following, we will formulate and analyze the AF design problems based on the above power constraints 1)-3).% and the constraint in 4) will be discussed as a remark after the theorems.

\subsection{A Unified AF Design Problem}

\begin{figure}[h]
  \centering
  \includegraphics[width=0.45\textwidth]{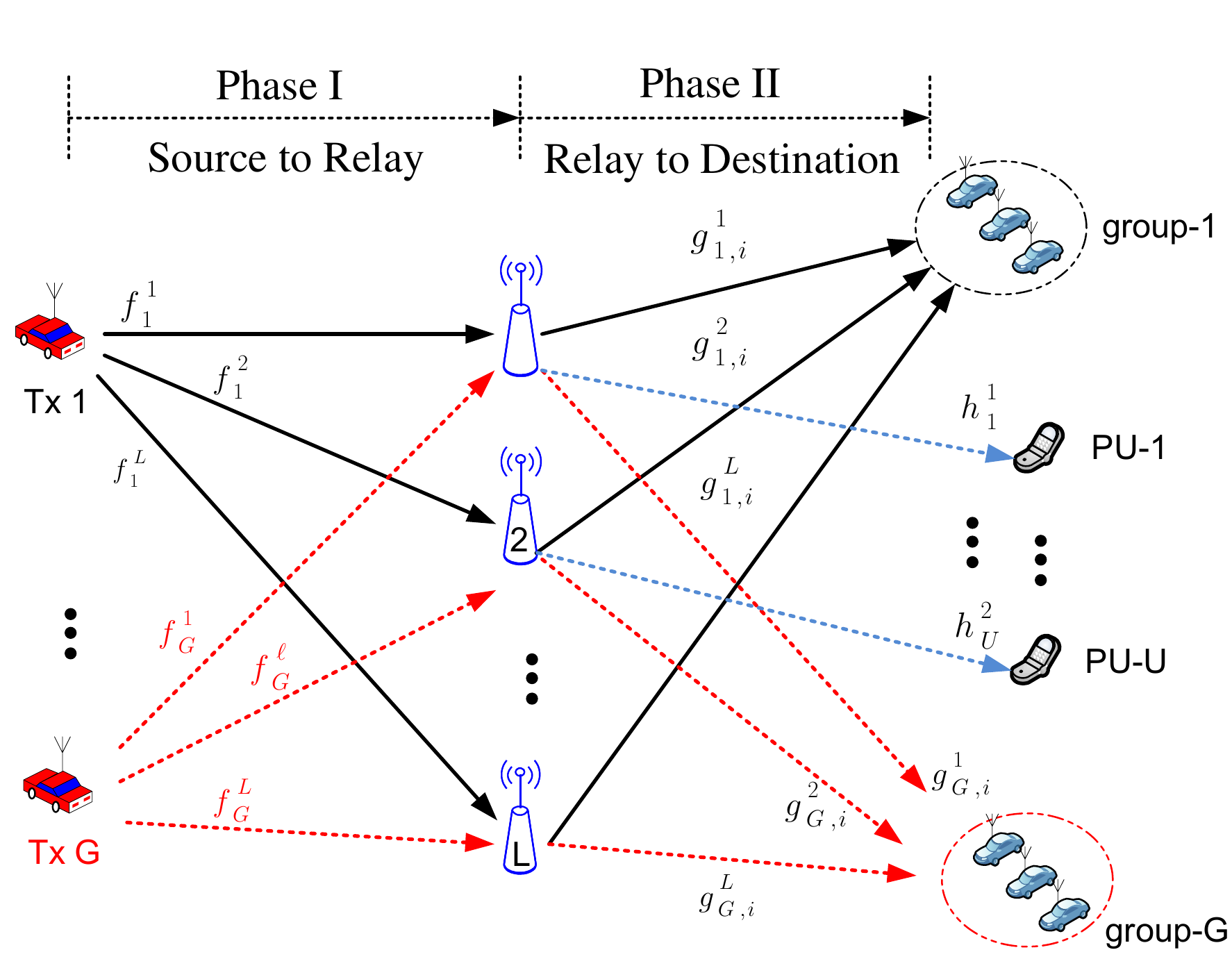}
\caption{The two-hop one-way relay network with primal users.}
\label{fig:2hop_relay_MM}
\end{figure}

In this sub-section, as depicted in Fig.\ref{fig:2hop_relay_MM}, we show how the beamformer design for different relay networks can be captured by a unified formulation.
%As depicted in Fig.\ref{fig:2hop_relay_MM},  we assume that there are single-antenna relays distributively located in the network; several primal users are present and they are far-apart from the transmitters; the relays are required to control the AF process so that no excessive interference is caused towards the primal users. To formulate the AF design, it is matter of whether the relays interact with each other and how they interact. In this paper, we study two specific cases, i.e., the MIMO relay network and the distributed relay network as follows: 

\subsubsection{Problem Formulation for the MIMO CR Relay Network}
%If all the relays fully share received signals, it will give rise to the MIMO relay network \cite{Jnl:Relay_Chalise_09}.
In light of the SINR expression in \eqref{gamma}, we consider the MMF design approach, which maximizes the worst user's SINR, subject to the power constraints. This gives rise to a one-variable fractional QCQP Problem (R1BF) in Table \ref{tab:BFAF_summary}. 
Note that we re-define $p_\ell$ with $p_0 = \bar P_0$ for $\ell=0$ and $p_\ell = \bar P_\ell$ for $\ell=1,...,L$.

\subsubsection{Problem Formulation for the Distributed CR Relay Network}
As for the distributed relays, since the AF weighting matrix here is a diagonal matrix, by a slight abuse of notation, we may re-define 
${\bm w}= {\rm Diag} ({\bm V})$~\footnote{Note that ${\bm V}$ is a diagonal matrix in the distributed relay network.}
and then the design problem for the distributed CR relay network can be written as exactly the same form as that for the MIMO relay, i.e., Problem (R1BF), given that $\mathcal{L}$, ${\bm A}_{k,i}$, ${\bm C}_{k,i}$, ${\bm D}_\ell$ and ${\bm G}_u$ are defined in \eqref{L}, \eqref{ak}, \eqref{ck}, \eqref{d0}, \eqref{dell}, and \eqref{Gu} in Table \ref{tab:BFAF_summary}. Note that herein the beamformer ${\bm w}$ is defined differently from that in the MIMO relay network. 

\subsubsection{Problem Formulation for Other One-hop Relay Networks}
Given that the way of interaction between relays is different, we may formulate different AF problems for design purposes. In fact, we could define ${\bm w} = {\rm vec}({\bm V}) \in \mathcal{V}$ where $\mathcal{V}$ specifies the prescribed possible structure of the AF matrix. One possible scenario is that the relays are partitioned into several groups, and the ones within the same group can fully communicate while the inter-group message passing is with limited capacity. This is typical when the backhaul is not powerful enough to support a massive information sharing. In this paper, we only focus on the MIMO relay network and the distributed relay network and provide a comprehensive performance analysis for them.

\ifconfver
    \begin{table*}[th]
\else
    \begin{table}[h]
\fi
\caption{Summary of the Design Problem for BF AF Schemes}
\label{tab:BFAF_summary}
\linespread{1.25} \rm \footnotesize
%\hline
\begin{tabular}{c||c}
~\\
\begin{minipage}{0.45\textwidth}
{
\begin{align*}
({\rm R1BF}) & \quad {\bm w}^\star =   \displaystyle \arg \max_{{\bm w} \in \mathbb{C}^{\mathcal{L}}}\min_{k=1,\ldots,G; \atop i=1,...,m_k} \frac{{\bm w}^H{\bm A}_{k,i}{\bm w}} {{\bm w}^H{\bm C}_{k,i}{\bm w}+1} \\
 \text{subject to} &\quad {\bm w}^H {\bm D}_\ell{\bm w}\le  \bar P_\ell, \quad \forall \ell=0,1,...,L, \\\notag
&\quad {\bm w}^H {\bm G}_u{\bm w}\le  \eta_u, \quad \forall u=1,...,U.
%\\&\quad {\bm w}^H {\bm R}_s {\bm w} \ge \rho_s, \quad \forall s=1,...,S,
\end{align*}
}
\end{minipage}
&
\begin{minipage}{0.45\textwidth}
{
\begin{align*}
{(\rm R1SDR)} &\quad {\bm W}^\star =  \displaystyle \arg\max_{{\bm W} \in \mathbb{H}_+^{{\mathcal L}}}\frac{{\bm A}_{k,i}\bullet {\bm W}} {{\bm C}_{k,i}\bullet{\bm W}+1} \\
\text{subject to}  & \quad {\bm D}_\ell \bullet {\bm W} \le \bar P_\ell, \quad \forall \ell=0,1,...,L, \\\notag
&\quad {\bm G}_u\bullet {\bm W}\le  \eta_u, \quad \forall u=1,...,U.
%\\
%&\quad {\bm R}_s \bullet {\bm W} \ge \rho_s, \quad \forall s=1,...,S,
\end{align*}
}
\end{minipage}
\\ 
\hline
~\\
\begin{minipage}{0.45\textwidth}
MIMO relay: \\
{
\begin{align}\label{L_v}
{\mathcal L} =& L^2 \\\label{ak_v}
{\bm A}_{k,i} =& P_k({\bm f}_k^* \otimes {\bm g}_{k,i})({\bm f}_k^* \otimes {\bm g}_{k,i})^H/\sigma_{k,i}^2 \\ \label{ck_v}
{\bm C}_{k,i} =& \displaystyle \sum_{m \neq k}P_m({\bm f}_m^* \otimes {\bm g}_{k,i})({\bm f}_m^* \otimes {\bm g}_{k,i})^H/\sigma_{k,i}^2 \\\notag
& ~~~~~+ {\bm \Sigma} \otimes ({\bm g}_{k,i}{\bm g}_{k,i})^H/{\sigma_{k,i}^2} \\\label{d_v0}
{\bm D}_0  =& \left(\sum_{j=1}^GP_j({\bm f}_j^*)({\bm f}_j^*)^H + {\bm \Sigma}\right) \otimes {\bm I} \\\label{d_v}
{\bm D}_\ell  =& \left(\sum_{j=1}^GP_j({\bm f}_j^*)({\bm f}_j^*)^H + {\bm \Sigma}\right) \otimes ({\bf e}_{\ell}{\bf e}_{\ell}^H)%~~~~\ell=1,...,L 
\\\label{G_u} 
{\bm G}_u  =& \displaystyle \sum_{j=1}^G P_j({\bm f}_j^* \otimes {\bm h}_{u})({\bm f}_j^* \otimes {\bm h}_{u})^H/\sigma_{u}^2
%\\\label{R_s}
%{\bm R}_s  =& \displaystyle \sum_{j=1}^G P_j({\bm f}_j \otimes ({\bm r}_{s})^*)({\bm f}_j \otimes ({\bm r}_{s})^*)^H
\end{align}
}
\end{minipage}
& 
\begin{minipage}{0.45\textwidth}
Distributed relay:\\
{
\begin{align} \label{L}
{\mathcal L} = & L \\\label{ak}
{\bm A}_{k,i} = &P_k({\bm f}_k^*\odot {\bm g}_{k,i})({\bm f}_k^*\odot {\bm g}_{k,i})^H/\sigma_{k,i}^2 \\ \label{ck}
{\bm C}_{k,i} = &\sum_{m \neq k}P_m({\bm f}_m^*\odot {\bm g}_{k,i})({\bm f}_m^*\odot {\bm g}_{k,i})^H/\sigma_{k,i}^2 \\ \notag
&+{\rm Diag} (|g_{k, i}^1|^2\sigma_1^2, |g_{k, i}^2|^2\sigma_2^2,...,|g_{k, i}^L|^2\sigma_L^2)/\sigma_{k,i}^2\\\label{d0}
{\bm D}_0 = &\sum_{j=1}^GP_j{\rm Diag}(({\bm f}_j^*)({\bm f}_j^*)^H) + {\bm \Sigma}\\\label{dell}
{\bm D}_\ell = &\left(\sum_{j=1}^GP_j{\rm Diag}(({\bm f}_j^*)({\bm f}_j^*)^H) + {\bm \Sigma}\right)\odot ({\bf e}_{\ell}{\bf e}_{\ell}^H)%~~~\ell=1,...,L 
\\\label{Gu}
{\bm G}_u  = & \displaystyle \sum_{j=1}^G P_j({\bm f}_j^*\odot {\bm h}_{u})({\bm f}_j^* \odot {\bm h}_{u})^H/\sigma_{u}^2
%\\\label{Rs}
%{\bm R}_s  =& \displaystyle \sum_{j=1}^G P_j({\bm f}_j \odot ({\bm r}_{s})^*)({\bm f}_j \odot ({\bm r}_{s})^*)^H.
\end{align}
}
\end{minipage}
\end{tabular}
\\  \hrule
\ifconfver
    \end{table*}
\else
    \end{table}
\fi

\subsection{The SDR Technique and Approximation Bounds for One-variable QCQPs}
From preceding discussions,  a generalized fractional QCQP problem (R1BF) is formulated for both the MIMO relay network and the distributed relay network. 
%This problem can be solved in the decision center (e.g., the cloud center) and an optimized ${\bm w}$ is obtained to guide the AF process.
%{\footnote{Note that ${\bm w}$ has different definitions in the MIMO relay network and the distributed relay network.}}
It is well known that Problem (R1BF) is in general NP-hard \cite{fazeli2009multiple, jimulti13, Jnl:MagzineMaLUO, chang2008approximation}. A classic way to tackle this NP-hard problem is to apply the SDR technique \cite{Jnl:MagzineMaLUO}, which reformulates QCQP as a rank constrained SDP problem and then drops the non-convex rank constraint. Specifically, to put SDR into the context, we use the equivalence
%\begin{equation} \label{eq:rank-eqv}
${\bm W} = {\bm w}{\bm w}^H \,\Longleftrightarrow\, {\bm W} \succeq {\bm 0}, \, \mbox{rank}({\bm W}) \le 1$
%\end{equation}
to relax (R1BF) to (R1SDR) as shown in Table \ref{tab:BFAF_summary}. It is well known that (R1SDR) can be approximated to arbitrary accuracy by the bisection method \cite{chang2008approximation, jimulti13}, where at each iteration of the bisection, an SDP problem is solved in polynomial time \cite{Jnl:Karipidis_MM_2008}. Let us define
$$%\begin{align}\label{gamma}
\gamma(\mathbf{W}) =  \min_{k=1,\ldots,G; \atop i=1,...,m_k} \frac{{\bm A}_{k,i}\bullet {\bm W}} {{\bm C}_{k,i}\bullet{\bm W}+1}
$$%\end{align} 
and $\bm{W}^\star$ and ${\bm{w}^\star}$ be the optimal solutions to (R1SDR) and (R1BF), respectively.
It follows that
$
\gamma(\bm{W}^\star) \ge \gamma({\bm{w}^\star}(\bm{w}^\star)^H),
$
since (R1SDR) is a convex relaxation of (R1BF).
The equality holds when (R1SDR) has rank-one solutions. If ${\rm rank}(\bm{W}^\star) > 1$, a feasible but generally sub-optimal solution $\widehat{\bm w}$ is extracted from ${\bm{W}}^\star$ by a Gaussian randomization algorithm \cite{MulticastLuo07, chang2008approximation, jimulti13}; more specifically, see Algorithm 1 in \cite{wu2015relaysbf}. It is easy to see that
$
 \gamma({\bm{w}^\star}(\bm{w}^\star)^H) \ge \gamma(\widehat{\bm w}(\widehat{\bm w})^H).
$
We are now interested in determining whether a reverse inequality (approximately) holds, as such an inequality will reveal the approximation quality of the SDR solution.  We summarize the prior results in \cite{wu2015relaysbf, chang2008approximation} as follows.
\begin{Prop} \quad
\label{Prop:main}
Let $M\ge1$ be the total number of users in the relay network and $J$ denote the total number of constraints in (R1BF), i.e., $J=L+U+1$ and $J \ge 1$. Then,
\begin{itemize}
\item[a).]When $ M+J \le 4$, (R1SDR) always produces an optimal solution to (R1BF).
\item[b).]When $M + J > 4$, let $\widehat{\bm w}$ be the solution returned by Gaussian randomization algorithm and $N$ be the number of randomizations. Then, with probability at least $1-(5/6)^N$, we have
\[
\gamma\left( \widehat{\bm w} \widehat{\bm w}^H\right) = \Omega\left(\frac{1}{M\log J}\right) \gamma(\bm{w}^\star{\bm{w}^\star}^H). 
\]
\end{itemize}
\end{Prop}
\noindent
Proposition \ref{Prop:main} implies that the SDR-based BF AF scheme is optimal when $M+J \le 4$. However, it will experience an SINR performance degradation when $M$ or $J$ is large. Specifically, in the worst case, the approximate quality is on the order of $1/(M\log J)$. Therefore, it may not work well in large-scale systems. 

\noindent
{\it Remark 1:}
Recently, there is flourishing interest in using RF wave to transfer power to those users in the system who aim at receiving energy rather than information. We usually call this kind of users as \emph{energy receivers} (ERs). As in the case with primal users, we could impose $R$ harvested energy constraints to the design problem \cite{Varshney,Grover,Zhang}. In this case, Proposition 1a) still holds by setting $J=L+U+R+1$. However, a similar result to Proposition 1b) holds only when $R=1$. This statement can be proven in a similar manner as Proposition 1 by using Theorem 2 in \cite{he2008semidefinite}.  

\section{The Rank-Two Beamformed Alamouti AF Scheme in the Relay Networks}
Our discussions in Section II reveal that BF AF may experience serious performance degradation in large-scale systems. 
However, in practice, the multigroup multicast transmission aims to serve groups of users and also, each of the relay should have its own power budget, not to mention that there may be a number of primal users if we consider the CR system. Therefore, the inherent feature of the relay network motivates us to propose more advanced AF schemes, for which the guaranteed performance can scale better with the problem size. We observe in Proposition \ref{Prop:main} that the performance degradation results from using  a rank-one AF weight ${\bm w}$ to approximate a high rank SDP solution ${\bm W}^\star$. To remedy this defect, an intuition is to explore one more DoF by introducing the Alamouti space-time code. This motivates the development of the BFA AF scheme.

\subsection{The Beamformed Alamouti AF Scheme for the MIMO relay}

\begin{figure}[h]
  \centering
  \includegraphics[width=0.42\textwidth]{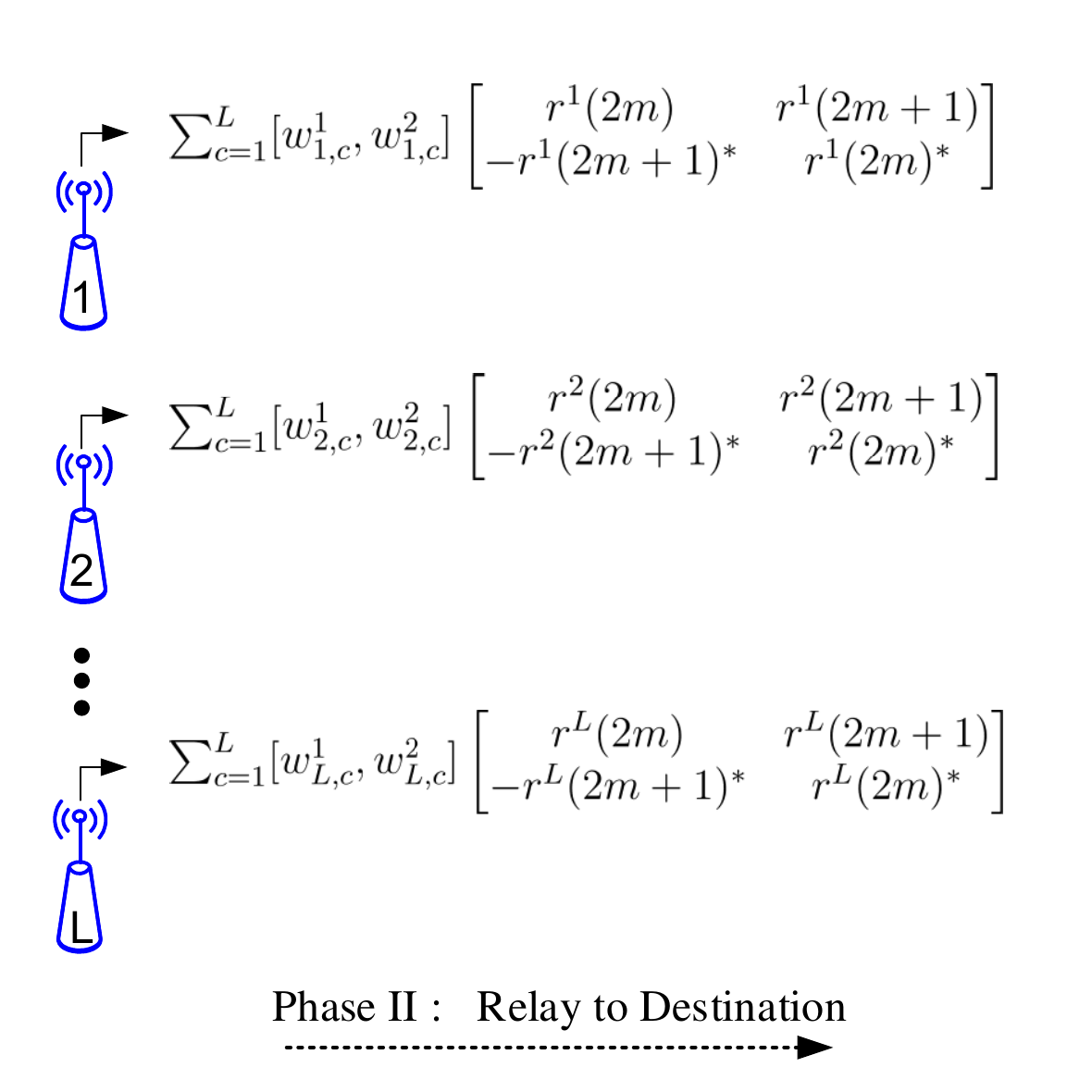}
\caption{The BFA AF signal structure for the MIMO relay.}
\label{fig:rank_two_tx_mimo}
\end{figure}

To introduce the Alamouti space-time code structure in the AF process, we parse the transmit signal in every two time slots as a group, i.e.,  ${\bf s}(m)=[~ s(2m) \,\,\, s(2m+1) ~]^T$, and transmit ${\bf s}(m)$ in each source-to-relay hop. Specifically, following \eqref{eq:xt_v}, for any $p=1,2$, we denote the AF weighting matrix at time slot $p$ as ${\bm V}_p = [{\bm v}_{1}^p, ...,{\bm v}_{\ell}^p,...{\bm v}_{L}^p]$ where ${\bm v}_{\ell}^p = [v_{1, \ell}^{p},..., v_{\ell, \ell}^{p},...,v_{L, \ell}^{p}]^T$. 
Then,
the transmit signal at relay $\ell$ (for every two time slots) can be modified to
\begin{align*}%\label{Xm}
{\bm X}_\ell(m) = &[~ {X}_\ell(2m) \,\,\, {X}_\ell(2m+1) ~] \\
=&\sum_{c=1}^L[v_{\ell, c}^1, v_{\ell, c}^2]\mathbf{C}( {\bf r}^{c}(m) ),
\end{align*}
where $\mathbf{C}: \mathbb{C}^2 \rightarrow \mathbb{C}^{2 \times 2}$ is the Alamouti space-time code, e.g.,
$ {\bf C}({\bm x}) = \begin{bmatrix} x_1 & x_2 \\ -x_2^* & x_1^* \end{bmatrix}, $
and we denote ${\mathbf{r}}^\ell(m)= [~ r^\ell(2m) \,\,\, r^\ell(2m+1) ~]^T$ with $r^\ell(2m)$ defined in \eqref{rt_v}. To demonstrate the signal structure, 
we illustrate the transmit signal at each relay in Figure~\ref{fig:rank_two_tx_mimo}.
The corresponding receive signal at user-$(k, i)$ is given in \eqref{chapter2:eq:model_alam_uni_relay_mimo}.
\begin{align}\label{chapter2:eq:model_alam_uni_relay_mimo}
\quad&{\bf y}_{k, i}(m) \quad \\\notag
=\quad& [~ y_{k, i}(2m), ~ y_{k, i}(2m+1) ~]\\\notag
\quad=\quad  &\sum_{\ell=1}^L({{g}_{k, i}^\ell})^*\sum_{c=1}^{L}[v_{\ell, c}^1, v_{\ell, c}^2]\mathbf{C}( {\bf r}^{c}(m) ) + [{{\mu}_{k, i}}(2m), ~ {{\mu}_{k, i}}(2m+1)]
,\\ \notag
\quad=\quad  &\sum_{\ell=1}^L({{g}_k^\ell})^*\sum_{c=1}^{L}[v_{\ell, c}^1, v_{\ell, c}^2] \begin{bmatrix} r^c(2m) & r^c(2m+1) \\ -r^c(2m+1)^* & r^c(2m)^* \end{bmatrix} + [{{\mu}_{k, i}}(2m), ~ {{\mu}_{k, i}}(2m+1)], \\ \notag
\quad=\quad  & \underbrace{\sum_{\ell=1}^L\sum_{c=1}^{L}[({{g}_{k, i}^\ell})^*v_{\ell, c}^1{f}_k^c, ({{g}_{k, i}^\ell})^*v_{\ell, c}^2{{f}_k^c}^*]  \begin{bmatrix}  s_k(2m) &  s_k(2m+1) \\ - s_k(2m+1)^* &  s_k(2m)^* \end{bmatrix}}_{\rm desired~signal} \\ \notag
&+\quad\underbrace{\sum_{\ell=1}^L\sum_{c=1}^{L}\sum_{j \neq k}[({{g}_{k, i}^\ell})^*v_{\ell, c}^1{f}_j^c, ({{g}_{k, i}^\ell})^*v_{\ell, c}^2{{f}_j^c}^*]  \begin{bmatrix}  s_j(2m) &  s_j(2m+1) \\ - s_j(2m+1)^* &  s_j(2m)^* \end{bmatrix}}_{\rm interference~signal} \\ \notag
&+\underbrace{\sum_{\ell=1}^L\sum_{c=1}^{L}[({{g}_{k, i}^\ell})^*v_{\ell, c}^1, ({{g}_{k, i}^\ell})^*v_{\ell, c}^2]  \begin{bmatrix}  {n}^\ell(2m) &  {n}^\ell(2m+1) \\ - {n}^\ell(2m+1)^* &  {n}^\ell(2m)^* \end{bmatrix} + [{{\mu}_{k, i}}(2m), ~ {{\mu}_{k, i}}(2m+1)]}_{\rm noise}.
\end{align}
%\begin{figure*}[!btp]
%\normalsize
%\begin{align}\label{chapter2:eq:model_alam_uni_relay_distri}
%&{\bf y}_{k, i}(m) \quad = [~ y_{k, i}(2m), ~ y_{k, i}(2m+1) ~] \\ \notag
%=&\sum_{\ell=1}^L({{g}_{k, i}^\ell})^*[w_{1, \ell}, w_{2, \ell}]\mathbf{C}( {\bf r}_{\ell}(m) ) + [{{v}_{k, i}}(2m), ~ {{v}_{k, i}}(2m+1)]
%,\\ \notag
%\quad=\quad  &\sum_{\ell=1}^L({{g}_k^\ell})^*[w_{1, \ell}, w_{2, \ell}]  \begin{bmatrix} r^\ell(2m) & r^\ell(2m+1) \\ -r^\ell(2m+1)^* & r^\ell(2m)^* \end{bmatrix} + [{{v}_{k, i}}(2m), ~ {{v}_{k, i}}(2m+1)], \\ \notag
%\quad=\quad  &\sum_{\ell=1}^L[({{g}_{k, i}^\ell})^*w_{1, \ell}{f}_k^\ell, ({{g}_{k, i}^\ell})^*w_{2, \ell}{{f}_k^\ell}^*]  \begin{bmatrix}  s_k(2m) &  s_k(2m+1) \\ - s_k(2m+1)^* &  s_k(2m)^* \end{bmatrix} \\ \notag
  %\quad &+\sum_{\ell=1}^L[({{g}_{k, i}^\ell})^*w_{1, \ell}, ({{g}_{k, i}^\ell})^*w_{2, \ell}]  \begin{bmatrix}  {n}^\ell(2m) &  {n}^\ell(2m+1) \\ - {n}^\ell(2m+1)^* &  {n}^\ell(2m)^* \end{bmatrix} + [{{v}_{k, i}}(2m), ~ {{v}_{k, i}}(2m+1)].
%\end{align}
%\hrulefill
%\vspace*{4pt}
%\end{figure*}
By using a similar approach in identifying the signal-to-noise ratio (SNR)  for the Alamouti coded signal in the point-to-point communication \cite[Chapter 3.3.2]{tse2005fundamentals}, we can derive the SINR expression from \eqref{chapter2:eq:model_alam_uni_relay_mimo} by treating interference as white noise. Specifically, letting ${\bm w}_1 = {\rm vec}({\bm V}_1)$ and ${\bm w}_2 = {\rm vec}({\bm V}_2)$,
the SINR at user-$(k, i)$ can be written as %\footnote{We measure the SINR of every two consecutive symbols in the same Alamouti code block.}
\begin{align}
\frac{{\bm w}_1^H{\bm A}_{k,i}{\bm w}_1 + {\bm w}_2^H\bar{\bm A}_{k,i}{\bm w}_2}{{\bm w}_1^H{\bm C}_{k,i}{\bm w}_1 + {\bm w}_2^H\bar{\bm C}_{k,i}{\bm w}_2+1}, 
\end{align}
where ${\bm A}_{k,i}$, $\bar{\bm A}_{k,i}$, ${\bm C}_{k,i}$ and $\bar{\bm C}_{k,i}$ are defined in \eqref{ak_v}, \eqref{ak_v_bar}, \eqref{ck_v} and \eqref{ck_v_bar}.  More details of the BFA AF detecting process can be found in the companion technical report~\cite{CompTechRepr2}. Similarly, we can obtain $\bar{\bm D}_\ell$ and $\bar{\bm G}_u$ in \eqref{d_v_bar} and \eqref{G_u_bar}.

\subsection{The Beamformed Alamouti AF Scheme for the distributed relay}
The BFA AF scheme for the distributed relay is similar to that for the MIMO case, and we defer the details of the derivations to the 
the companion technical report~\cite{CompTechRepr2}. Note that for distributed relay network, we re-define ${\bm w}_1= {\rm Diag} ({\bm V}_1)$ and ${\bm w}_2= {\rm Diag} ({\bm V}_2)$, which is different from that in MIMO relay network.
Therefore, the design problem for both the MIMO relay and distributed relay can be written as a two-variable fractional QCQP problem (R2BF) in Table \ref{tab:BFALAAF_summary}.

\ifconfver
    \begin{table*}[th]
\else
    \begin{table}[h]
\fi
\caption{Summary of the Design Problem for BF Alamouti AF Schemes}
\label{tab:BFALAAF_summary}
\linespread{1.25} \rm \footnotesize
%\hline
\begin{tabular}{c||c}
~\\
\begin{minipage}{0.45\textwidth}
{
\begin{align*}
&({\rm R2BF})\quad({\bm w}_1^\star, {\bm w}_2^\star) \\
&=\displaystyle{\arg\max}_{{\bm w}_1, {\bm w}_2 \in \mathbb{C}^\mathcal{L}}\displaystyle \displaystyle\min_{{i=1,\ldots,m_k \atop k=1,\ldots,G}} \frac{{\bm w}_1^H{\bm A}_{k,i}{\bm w}_1^H + {\bm w}_2^H\bar{\bm A}_{k,i}{\bm w}_2}{{\bm w}_1^H{\bm C}_{k,i}{\bm w}_1 + {\bm w}_2^H\bar{\bm C}_{k,i}{\bm w}_2+1} \\
&\text{subject to}\quad {\bm w}_1^H {\bm D}_\ell{\bm w}_1 + {\bm w}_2^H\bar{\bm D}_\ell{\bm w}_2 \le  \bar P_\ell, \quad \forall \ell=0,1,...,L, \\\notag
&\quad\quad \quad\quad \quad   {\bm w}_1^H {\bm G}_u{\bm w}_1 + {\bm w}_2^H \bar{\bm G}_u{\bm w}_2 \le  \eta_u, \quad \forall u=1,...,U.
\end{align*}
}
\end{minipage}
&
\begin{minipage}{0.45\textwidth}
{
\begin{align*}
&({\rm R2SDR}) \quad ({\bm W}_1^\star, {\bm W}_2^\star) \\
&=\displaystyle{\arg\max}_{{\bm W}_1, {\bm W}_2 \in \mathbb{H}_+^{\mathcal{L}}} \displaystyle\min_{{i=1,\ldots,m_k \atop k=1,\ldots,G}} \frac{{\bm A}_{k,i}\bullet{\bm W}_1 + \bar{\bm A}_{k,i}\bullet{\bm W}_2}{{\bm C}_{k,i}\bullet {\bm W}_1 + \bar{\bm C}_{k,i}\bullet{\bm W}_2 +1}\\
& \text{subject to} \quad {\bm D}_\ell \bullet {\bm W}_1 + \bar{\bm D}_\ell\bullet{\bm W}_2 \le  \bar P_\ell, \quad \forall \ell=0,1,...,L, \\\notag
&\quad \quad \quad \quad \quad {\bm G}_u\bullet{\bm W}_1 + \bar{\bm G}_u\bullet{\bm W}_2 \le  \eta_u, \quad \forall u=1,...,U.
\end{align*}
}
\end{minipage}
\\ 
\hline
~\\
\begin{minipage}{0.45\textwidth}
MIMO relay: \\
{
\begin{align}\notag
{\mathcal L} =& L^2 \\\label{ak_v_bar}
\bar {\bm A}_{k,i} & = P_k({\bm f}_k \otimes {\bm g}_{k,i})({\bm f}_k \otimes {\bm g}_{k,i})^H/\sigma_{k,i}^2 \\\notag
\bar {\bm C}_{k,i} & = \displaystyle \sum_{m \neq k}P_m({\bm f}_m \otimes {\bm g}_{k,i})({\bm f}_m \otimes {\bm g}_{k,i})^H/\sigma_{k,i}^2 \\\label{ck_v_bar}
& ~~~~~+ {\bm \Sigma} \otimes ({\bm g}_{k,i}{\bm g}_{k,i})^H)/{\sigma_{v,i}^2}, \\\label{d_v_bar}
\bar{\bm D}_\ell  &= \left(\sum_{j=1}^GP_j{\bm f}_j{\bm f}_j^H + {\bm \Sigma}\right) \otimes ({\bf e}_{\ell}{\bf e}_{\ell}^H)\\\label{G_u_bar}
\bar{\bm G}_u &= \displaystyle \sum_{j=1}^G P_j({\bm f}_j \otimes {\bm h}_{u})({\bm f}_j \otimes {\bm h}_{u})^H/\sigma_{u}^2
%\\\label{R_s_bar}
%{\bm R}_s  &= \displaystyle \sum_{j=1}^G P_j(({\bm f}_j)^* \otimes ({\bm r}_{s})^*)(({\bm f}_j)^* \otimes ({\bm r}_{s})^*)^H
\end{align}
}
\end{minipage}
& 
\begin{minipage}{0.45\textwidth}
Distributed relay:\\
{
\begin{align} \notag
{\mathcal L} = & L \\\label{ak_bar}
\bar {\bm A}_{k, i}  = & P_k({\bm f}_k\odot {\bm g}_{k,i})({\bm f}_k\odot {\bm g}_{k,i})^H/\sigma_{k,i}^2 \\\notag
\bar {\bm C}_{k,i}  = & \displaystyle\sum_{m \neq k}P_m({\bm f}_m\odot {\bm g}_{k,i})({\bm f}_m\odot {\bm g}_{k,i})^H/\sigma_{k,i}^2 \\\label{ck_bar}
&+{\rm Diag} (|g_{k, i}^1|^2\sigma_1^2, |g_{k, i}^2|^2\sigma_2^2,...,|g_{k, i}^L|^2\sigma_L^2)/\sigma_{k,i}^2, \\\label{dv_bar}
\bar{\bm D}_\ell  =&\left(\sum_{j=1}^GP_j{\rm Diag}({\bm f}_j{\bm f}_j^H) + {\bm \Sigma}\right)\odot ({\bf e}_{\ell}{\bf e}_{\ell}^H)\\\label{Gu_bar}
\bar {\bm G}_u  = & \displaystyle \sum_{j=1}^G P_j({\bm f}_j \odot {\bm h}_{u})({\bm f}_j \odot {\bm h}_{u})^H/\sigma_{u}^2
%\\\label{Rs_bar}
%{\bm R}_s  &= \displaystyle \sum_{j=1}^G P_j(({\bm f}_j)^* \odot ({\bm r}_{s})^*)(({\bm f}_j)^* \odot ({\bm r}_{s})^*)^H
\end{align}
}
\end{minipage}
\end{tabular}
\\ \hrule
\ifconfver
    \end{table*}
\else
    \end{table}
\fi

\subsection{The SDR Technique for (R2BF)}

Problem (R2BF) is a fractional QCQP with two variables. It subsumes Problem (R1BF) when ${\bar {\bm A} = {\bm 0}}$, ${\bar {\bm C} = {\bm 0}}$ and ${\bar {\bm G} = {\bm 0}}$, and thus it is NP-hard. By applying SDR, we have the new AF design problem shown in
%We 
%can apply 
%$
%%\begin{equation} \label{eq:rank-eqv_v}
%{\bm W} = {\bm w}{\bm w}^H \,\Longleftrightarrow\, {\bm W} \succeq {\bm 0}, \, \mbox{rank}({\bm W}) \le 2,
%%\end{equation}
%$
%and drop he rank constraint. 
Table \ref{tab:BFALAAF_summary}, where
(R2SDR) serves as a convex relaxation to (R2BF). Define
\begin{equation}
\theta({\bm W}_1, {\bm W}_2) = \displaystyle\min_{{i=1,\ldots,m_k \atop k=1,\ldots,G}} \frac{{\bm A}_{k,i}\bullet{\bm W}_1 + \bar{\bm A}_{k,i}\bullet{\bm W}_2}{{\bm C}_{k,i}\bullet {\bm W}_1 + \bar{\bm C}_{k,i}\bullet{\bm W}_2 +1}.
\end{equation}
Clearly, we have
\begin{equation}\label{eq:W_w}
\theta({\bm W}_1^\star, {\bm W}_2^\star) \ge \theta({\bm w}_1^\star{{\bm w}_1^\star}^H, {\bm w}_2^\star{{\bm w}_2^\star}^H),
\end{equation}
and equality holds whenever Problem (R2SDR) returns rank-one solutions. If ${\bm W}_1^\star$ or ${\bm W}_2^\star$ is non-rank-one,  the Gaussian randomization algorithm in Algorithm \ref{alg:1} is applied to generate a sub-optimal solution $(\hat{\bm w}_1,\hat{\bm w}_2)$ to (R2BF) with
$
\theta({\bm w}_1^\star{{\bm w}_1^\star}^H, {\bm w}_2^\star{{\bm w}_2^\star}^H) \ge \theta(\widehat{\bm w}_1{\widehat{\bm w}_1}^H, \widehat{\bm w}_2{\widehat{\bm w}_2}^H).
$
To the best of our knowledge, the approximation quality of the solution $(\hat{\bm w}_1,\hat{\bm w}_2)$ has not been addressed in the literature before.  In the next sub-section, we shall establish the first provable bound on the approximation quality of $(\hat{\bm w}_1,\hat{\bm w}_2)$.

\subsection{SDR Approximation Bounds for BF Alamouti AF}
In this sub-section, we turn our attention to the performance analysis of the proposed BFA AF scheme.  Specifically, we proceed by
answering the following three questions.

\begin{itemize}
\item[1).] What is the relationship between (R2SDR) and (R1SDR)?
\item[2).] When is (R2SDR) equivalent to (R2BF)?
\item[3).] What is the approximation quality of (R2SDR)?
\end{itemize}
Our recent works \cite{MainPaper, jimulti13} show that for the multigroup multicast network without relays, the design problems for BF AF and BFA AF have the same SDR, and the BFA AF scheme admits a rank-two approximate solution. However, the design problem for relay networks gives rise to 
two different SDRs, namely, (R1SDR) and (R2SDR).  To answer Problem 1), it is easy to see that 
\begin{Prop} \label{pro:r2betterr1}\quad
\label{Prop:main_feasible}
The optimal value of (R2SDR) is always no worse than that of (R1SDR).
\end{Prop}
\noindent
The proof comes from the fact that $(\widehat{\bm W}, {\bf 0})$ must be a feasible solution to (R2SDR), given that $\widehat {\bm W}$ is feasible to (R1SDR). Intuitively, the gain also comes from the employment of one more DoF in the transmit structure.
 
\begin{algorithm}[t]
\caption{Gaussian Randomization for (R2BF)} \label{alg:1}
\begin{algorithmic}[1]
\STATE input: optimal solutions $\bm{W}_1^{\star}, \bm{W}_2^{\star}$ to (R2SDR), number of randomizations $N \ge 1$
\FOR {$P=1$ to $2$}
\IF {$\mbox{rank }({\bm W}_p^\star) \le 1$}
\STATE let $\hat{\bm{w}}_p^n(\hat{\bm w}_p^n)^H = {\bm W}_p^\star$
\ENDIF
\ENDFOR
\FOR {$n=1$ to $N$}
\STATE for any $\mbox{rank}({\bm W}_p^\star) > 1$, we generate $\bm{\xi}_p^n \sim \mathcal{CN}(\bm{0}, \bm{W}_p^{\star})$ and let 
$\hat{\bm{w}}_p^n =  \bm{\xi}_p^n \bullet \min\left\{t_1, \quad t_2\right\}$
where
$
 t_1 = \min_{\ell}
\sqrt{\bar P_\ell/\left({\bm D}_\ell \bullet \left( {\bm{\xi}}_1^n {\bm{\xi}_1^n}^H\right) + \bar{\bm D}_\ell\bullet\left({\bm{\xi}}_2^n{\bm{\xi}_2^n}^H \right)\right)},
$ 
and
$
t_2 = \min_{u}
\sqrt{\eta_u/\left({\bm G}_\ell \bullet \left( \bm{\xi}_1^n {\bm{\xi}_1^n}^H \right)+ \bar{\bm G}_\ell \bullet \left( \bm{\xi}_2^n {\bm{\xi}_2^n}^H \right)\right)}
$
\STATE set $\theta_n =  \theta\left( \hat{\bm{w}}_1^n(\hat{\bm{w}}_1^n)^H, \hat{\bm{w}}_2^n(\hat{\bm{w}}_2^n)^H \right)$
\ENDFOR
\STATE set $n^\star = \arg\max_{n=1,\ldots,N} \theta_n$ and output $\widehat{\bm{w}}_1 = \hat{\bm{w}}_1^{n^\star}$ and $\widehat{\bm{w}}_2 = \widehat{\bm{w}}_2^{n^\star}$.
\end{algorithmic}
\end{algorithm}
To address Problem 2) and 3), we have the following theorem:
\begin{Theorem} \label{thm:main}
Consider Problem (R2SDR). Let $M\ge1$ be the total number of users in the relay network and $J$ denote the total number of constraints in (R1BF), i.e., $J=L+U+1$ and $J \ge 1$.  Let ${\bm W}_1^{\star}$ and ${\bm W}_2^{\star}$ be the optimal solutions to Problem (R2SDR) and we consider the cases of both ${\bm W}_1^{\star}\neq {\bm 0}$ and ${\bm W}_2^{\star}\neq {\bm 0}$.\footnote{It is a trivial case if ${\bm W}_1^{\star}={\bm W}_2^{\star}={\bm 0}$. If either ${\bm W}_1^{\star}={\bm 0}$ or ${\bm W}_2^{\star}={\bm 0}$, it reduces to the rank-one case and (R2SDR) is optimal for (R2BF) when $M + J \le 4$. } Then, we have the following results:
\begin{itemize}
\item[a).] 
Problem (R2SDR) can produce an optimal solution to Problem (R2BF) when $M + J \le 5$.
\item[b).] 
When $M + J > 5$ and ${\rm rank}({\bm W}_p^{\star}) > 1$ for some $p=1,2$ in (R2SDR), let
$\widehat{\bm w}_1, \widehat{\bm w}_2$ be the solutions returned by Algorithm \ref{alg:1}, and $J=L+U+1$ be the total number of power constraints.
Then, with probability at least $1-(7/8)^N$, we have
\begin{align*}%\label{c0}
&\theta( \widehat{\bm w}_1\widehat{\bm w}_1^H, \widehat{\bm w}_2\widehat{\bm w}_2^H ) \ge c \cdot \theta( {\bm w}_1^\star{{\bm w}_1^\star}^H, {\bm w}_2^\star{{\bm w}_2^\star}^H),
%\Omega\left(\frac{\theta( ({\bm w}_1^\star)({{\bm w}_1^\star})^H, ({\bm w}_2^\star)({{\bm w}_2^\star})^H)}{\min\{M\log J, ~\omega\sqrt{M}J\}}\right),
\end{align*}
where
\begin{align}\label{c}
c = \frac{\max\left\{\frac{\omega}{7\sqrt{M}}, ~\frac{1}{8M}\right\}}{2\log(16J) + 1},
\end{align}
$\omega = \displaystyle\min_{k, i} \left\{\frac{\min\{{\bm A}_{k, i}\bullet{\bm W}_1^\star, ~\bar{\bm A}_{k, i}\bullet{\bm W}_2^\star\}}{{\bm A}_{k, i}\bullet{\bm W}_1^\star + \bar{\bm A}_{k, i}\bullet{\bm W}_2^\star} \right\}$, and $N$ is the number of randomizations.
\end{itemize}
\end{Theorem}
\noindent
The proof of Theorem \ref{thm:main}a) is similar to that for Theorem 1 in \cite{wu2015relaysbf}; see details in Appendix A in \cite{wu2015relaysbf} and \cite[Theorem 5.1]{Jnl:Yongwei_rank}.
We relegate the proof of Theorem \ref{thm:main}b) in Appendix \ref{Thm1} at the end of this paper. The significance of Theorem \ref{thm:main} is twofold.  First, it shows that (R2SDR) is optimal for BFA AF when $M + J \le 5$. Then, in conjunction with Proposition \ref{pro:r2betterr1}, the proposed method must be better than BF AF in the case of $M + J \le 5$. Second, if (R2SDR) does not have rank-one solutions, Theorem \ref{thm:main} specifies the quality of the SDR solution returned by the Gaussian randomization algorithm---i.e., on the order of $\frac{1}{\sqrt{M}\log J}$ if $\frac{\omega}{7\sqrt{M}}\ge \frac{1}{8M}$; on the order of $\frac{1}{M\log J}$ if $\frac{\omega}{7\sqrt{M}} < \frac{1}{8M}$.
%the SINR associated with the approximate solution scales with an order of $\max\{\Omega\left(\frac{1}{M\log J}\right), \Omega\left(\frac{\omega}{\sqrt{M}\log J}\right)\}$ of that associated with the optimal solution. 
If we fix the number of power constraints $J$ and increase the number of users in the system, then we see that $\Omega\left(1/\sqrt{M}\right)$ scales better than $\Omega\left(1/M\right)$.  Moreover, we see from~\eqref{c} that the approximation quality is lower bounded by $\Omega\left(1/M\right)$. Hence, in conjunction with Proposition \ref{pro:r2betterr1}, the performance of BFA AF is provably no worse than that of BF AF.

\noindent
{\it Remark 2:}
One may observe that (R2BF) can be written in the form of (R1BF) by letting ${\bm w} = [{\bm w}_1; {\bm w}_2]$ and $\tilde{\bm A}_{k, i} = [{\bm A}_{k, i}, {\bm 0}; {\bm 0}, \bar {\bm A}_{k, i}]$. Thus, it immediately admits the inferior bound in Proposition \ref{Prop:main}.  Moreover, if $\bar{\bm A}_{k, i}={\bm 0}$,  the approximation bound in Theorem \ref{thm:main} reduces to that in Proposition \ref{Prop:main}. Thus Theorem \ref{thm:main} can be seen as a generalization of Proposition \ref{Prop:main} from one-variable fractional QCQPs to two-variable fractional QCQPs. 

\noindent
{\it Remark 3:}
If we let ${\bar {\bm A}_{k, i}} = {\bm A}_{k, i}$ and ${\bar {\bm C}_{k, i}} = {\bm C}_{k, i}$, Problem (R2BF) will be reduced to 
Problem (R1BF) by replacing the rank-one variable ${\bm w}$ to a rank-two matrix ${\bm B}$. We have studied the latter in \cite{jimulti13} by considering the multigroup multicast transmission in a network without relays, where the SDR rank-two approximation bounds are provided based on Problem (R1SDR). Hence, our results in Theorem \ref{thm:main} also generalize the approximation bounds in \cite[Theorem 1]{jimulti13}.

%\noindent
%{\it Remark 4:}
%In Theorem \ref{thm:main}, the approximation bound not only depends on the problem size, but also on $\omega$. This may not be a good news for people who want a deterministic result. In practice, it is good to know how exactly these $2$-DoF help gain performance from the transmit structure.
%In the next section, we will address this problem by investigating two special cases. 

\section{Special Cases with Deterministic Approximation Bounds}
In this section, by considering some special cases of (R2BF), we will demonstrate how the two DoFs offered by the BFA AF scheme can lead to improved performance over the BF AF scheme. 
In particular, we consider special cases in both distributed relay network and MIMO relay network, where by exploiting the special structures of the problems, we can obtain various strengthenings of Theorem~\ref{thm:main}.

\subsection{Multicasting for Distributed Relay Networks}
For the distributed relay case, by observing the structure of ${\bm A}_{k,i}$ and $\bar {\bm A}_{k,i}$, we can define $\phi_\ell = \arg(f_k^\ell)$ and ${\bf e} = [e^{j 2\phi_1},...,e^{j2\phi_\ell},...,e^{j2\phi_L}]^T$ with $j=\sqrt{-1}$ such that
\begin{equation}\label{1}
{\bm w}_2^H \bar{\bm A}_{k,i}{\bm w}_2
= ({\bm w}_2 \odot {\bf e})^H {\bm A}_{k,i}({\bm w}_2 \odot {\bf e})
\end{equation}
and
\begin{equation}\label{2}
{\bm w}_2^H \bar{\bm G}_{u}{\bm w}_2
= ({\bm w}_2 \odot {\bf e})^H {\bm G}_{u}({\bm w}_2 \odot {\bf e}).
\end{equation}
Then, letting $\bar {\bm w}_2 = {\bm w}_2 \odot {\bf e}$, we actually have
\begin{align}\label{3}
 \bar{\bm A}_{k,i}\bullet {\bm w}_2{\bm w}_2^H = {\bm A}_{k,i}\bullet \bar{\bm w}_2 \bar{\bm w}_2^H
\end{align}
and
\begin{align}\label{4}
 \bar{\bm G}_{u}\bullet {\bm w}_2{\bm w}_2^H = {\bm G}_{u}\bullet \bar{\bm w}_2 \bar{\bm w}_2^H.
\end{align}
On the other hand, we also have
\begin{align}\label{5}
 & \bar{\bm C}_{k,i}\bullet \bar{\bm w}_2 \bar{\bm w}_2^H ={\bm C}_{k,i}\bullet{\bm w}_2{\bm w}_2^H,
\end{align}
since ${\bm C}_{k,i}$ must be a diagonal matrix in the target relay network. Moreover, for distributed relays, we have
$
\bar{\bm D}_{\ell} = {\bm D}_{\ell},  \quad \forall \ell=0,...,L,
$
and ${\bm D}_{\ell}$s are also diagonal.
Therefore
\begin{align}\label{6}
{\bm D}_{\ell}\bullet \bar{\bm w}_2 \bar{\bm w}_2^H = {\bm D}_{\ell}\bullet {\bm w}_2 {\bm w}_2^H, \quad \forall \ell=1,...,L,
\end{align}
for the diagonal matrix ${\bm D}_{\ell}$.
Thus, Problem (R2BF) can be rewritten as
\begin{align}\label{eq:relay_network_multicast_r2} 
\displaystyle\max_{{\bm w}_1, {\bm w}_2 \in \mathbb{C}^{\mathcal L}}&\displaystyle\min_{{i=1,\ldots,m_k \atop k=1,\ldots,G}}  \frac{{\bm w}_1^H {\bm A}_{k,i}{\bm w}_1 + \bar{\bm w}_2^H {\bm A}_{k,i}\bar{\bm w}_2}{{\bm w}_1^H {\bm C}_{k,i}{\bm w}_1 + \bar{\bm w}_2^H {\bm C}_{k,i}\bar{\bm w}_2 +1} \\\notag
\text{subject to} &\quad {\bm w}_1^H{\bm D}_\ell{\bm w}_1 + \bar{\bm w}_2^H{\bm D}_\ell\bar{\bm w}_2 \le \bar P_\ell,\quad \ell=0,1,...,L,\\\notag
&\quad {\bm w}_1^H{\bm G}_u{\bm w}_1 + \bar{\bm w}_2^H{\bm G}_u\bar{\bm w}_2 \le \eta_u,\quad u=0,1,...,U,
\end{align}
where we have ${\mathcal L} = L$.
Clearly, the resulting SDR of \eqref{eq:relay_network_multicast_r2} has the same optimal value as that of (R1SDR). Moreover, it is easy to see that we have $\omega = 1/2$ here since ${\bm A}_{k,i} = \bar{\bm A}_{k,i}$ for all $k, i$. Note that herein we have $k=1$ for multicasting. Thus, we may conclude
\begin{Prop} \quad
\label{Prop:mimomulticast0}
For distributed relay multicasting with generalized power constraints, Problem (R2BF) can be rewritten as Problem \eqref{eq:relay_network_multicast_r2}, and therefore the approximation bound 
\begin{align*}
\theta( \widehat{\bm w}_1\widehat{\bm w}_1^H, \widehat{\bm w}_2\widehat{\bm w}_2^H ) 
= \Omega\left(\frac{1}{\sqrt{M}\log J}\right)\theta( {\bm w}_1^\star {{\bm w}_1^\star}^H, {\bm w}_2^\star{{\bm w}_2^\star}^H)
\end{align*}
holds with probability at least $1-(7/8)^N$.
\end{Prop}
\noindent
This result can also be obtained by applying the approximation bound results in \cite{jimulti13} to (R1SDR). It is worth mentioning that prior to this proposition,  the authors in \cite{schad2012convex} have provided a similar result by introducing a phase rotation at the AF process under a total power constraint.  Nevertheless, our derivation here reveals how the two DoFs lead to a gain in performance. %That is, based on the same SDR problem, BFA AF admits a higher quality rank-two solution rather than rank one. That is the contribution from BFA AF. 

\subsection{Multicasting for MIMO Relay Networks} 
We also have some interesting results for the MIMO relay multicasting network. It shows that in this case, the AF scheme possesses some special structure, which leads to an approximation bound of $\Omega(\frac{1}{\sqrt{M}\log J})$. We summarize the result in Proposition \ref{Thm2}:

\begin{Prop} \quad
\label{Thm2}
For MIMO relay multicasting with generalized power constraints, Problem (R2BF) can be rewritten as Problem \eqref{eq:relay_network_multicast_r2} with proper parameter definitions for MIMO relays, and therefore the approximation bound 
\begin{align*}
\theta( \widehat{\bm w}_1\widehat{\bm w}_1^H, \widehat{\bm w}_2\widehat{\bm w}_2^H ) 
= \Omega\left(\frac{1}{\sqrt{M}\log J}\right)\theta( {\bm w}_1^\star {{\bm w}_1^\star}^H, {\bm w}_2^\star{{\bm w}_2^\star}^H)
\end{align*}
holds with probability at least $1-(7/8)^N$.
\end{Prop}
\noindent
The proof is similar to that in Proposition \ref{Prop:mimomulticast0}, except that we now define $\phi_\ell = \arg(f_k^\ell)$ and ${\bf e} = [e^{j 2\phi_1},...,e^{j2\phi_\ell},...,e^{j2\phi_L}]^T \otimes {\bm 1}$ such that
equations \eqref{1}-\eqref{6} hold for the MIMO relay case. This proposition reveals that for MIMO relay multicasting, the SDR approximation bound is on the order of $1/({\sqrt{M}\log J})$, which is definitely better than that for the BF AF counterpart. It is worth noting that the optimal values for (R1SDR) and (R2SDR) are the same, and thus the SINR performance is improved due to the better approximation quality. This demonstrates how the proposed BFA AF scheme benefits the system performance.

\section{Numerical Simulations}\label{sec:sim}
In this section, we provide numerical simulations to compare the performance of different AF schemes and demonstrate the superiority of the proposed BFA AF scheme. We show the numerical results for the distributed relay network in this section and defer the results for the MIMO relay network and the comparison with the feasible point pursuit (FPP) algorithm \cite{mehanna2015feasible, christopoulos2015multicast} in the 
the companion technical report~\cite{CompTechRepr2}.  We assume w.l.o.g. that each multicast group has an equal number of users (i.e., $m_k=M/G$ for $k=1,\ldots,G$).  The channels ${\bm f}_k, {\bm g}_{k,i}$, where $k=1,\ldots,G$ and $i=1,\ldots,m_k$, are identical independently distributed (i.i.d.) according to $\mathcal{CN}({\bm 0}, {\bm I})$.  The transmitted signal at each transmitter is with power $0$dB (i.e., $P_j=0$dB for $j=1,\ldots,G$). Each single-antenna relay has the same noise power (i.e., $\sigma_{\ell}^2=\sigma_{\sf ant}^2$, where $\ell=1,\ldots,L$), and all users have the same noise power (i.e., $\sigma_{k,i}^2=\sigma_{\sf user}^2$ for $k=1,\ldots,G$ and $i=1,\ldots,m_k$). We assume that $\sigma_{\sf ant}^2>0$ and $\sigma_{\sf user}^2>0$.  The total power threshold for all the relays is $\bar{P}_0$; the power threshold at $\ell$th relay is $\bar{P}_\ell$, where $\ell=1,\ldots,L$. For each AF scheme, $100$ channel realizations were averaged to get the plots, and the number of randomizations for generating BF AF weights and BFA weights is $1,000$.

\subsection{Worst User's SINR versus Total Power Threshold} \label{subsec:tot-pow}
\begin{figure}[htb]
\centering
\includegraphics[width=0.42\textwidth]{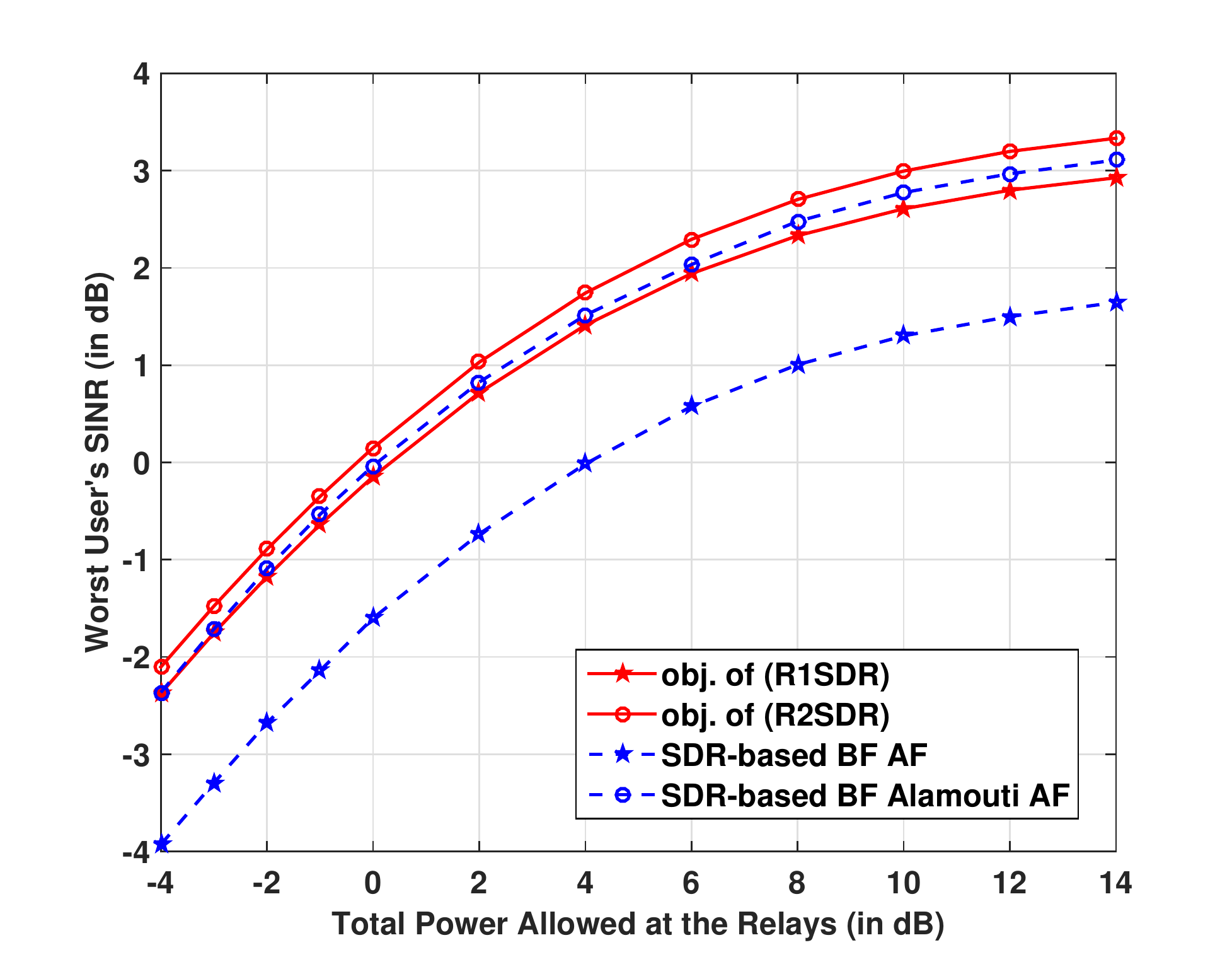}
\caption{Worst user's SINR versus total power threshold at the distributed relay: $L=8$, $G=2$, $M=12$, $\sigma_{\sf ant}^2=\sigma_{\sf user}^2=0.25$.}
\label{fig:2}
\end{figure}

In this simulation, we vary the total power budget at relays to see the worst user's SINR performance in different relay networks. For ease of exposition, 
we consider the scenario where only the total power constraint is present. In Figure \ref{fig:2}, we assume that there are $L=8$ single-antenna relays and $G=2$ multicast groups with a total of $M=16$ users. For both cases, we set $\sigma_{\sf ant}^2=\sigma_{\sf user}^2=0.25$. It shows that
the objective value (obj.) of (R1SDR) and (R2SDR) serve as an upper bound for the SDR-based BF AF scheme and the SDR-based BFA AF scheme, respectively. Moreover, based on randomization, the BFA AF scheme shows a significant SINR improvement over the BF AF scheme in all the power regions. 

%\begin{figure}[htb]
%\centering
%\includegraphics[width=0.38\textwidth]{./r2relaytsp/MIMO_MM_MMF_428_0db_025_025_100.eps}
%\caption{Worst user's SINR versus total power threshold at the MIMO relay: $L=4$, $G=2$, $M=16$, $\sigma_{\sf ant}^2=\sigma_{\sf user}^2=0.25$.}
%\label{fig:1}
%\end{figure}

\subsection{Worst User's SINR versus Number of Users}

\begin{figure}[htb]
\centering
\includegraphics[width=0.42\textwidth]{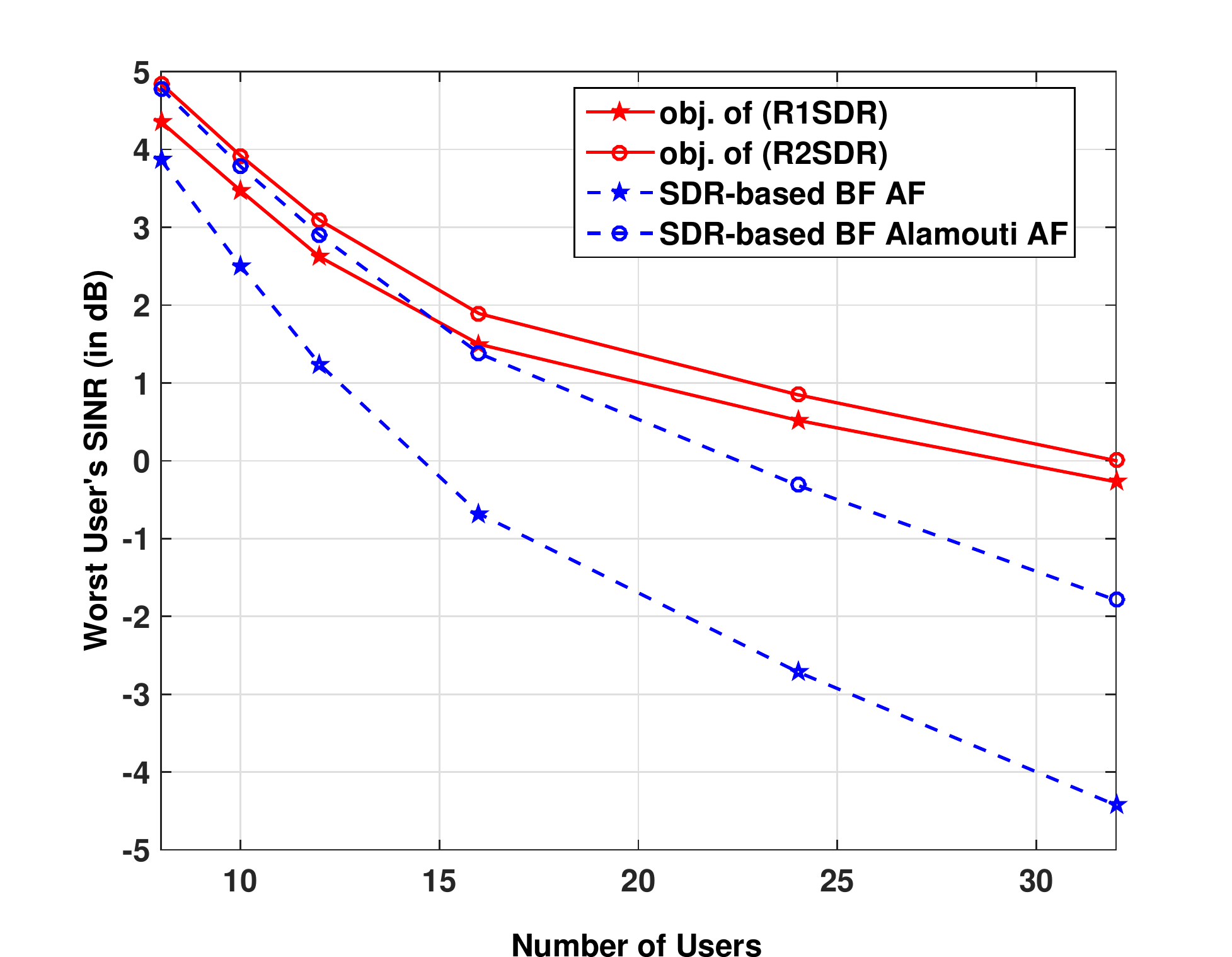}
\caption{Worst user's SINR versus number of users in the distributed relay system: $L=8$, $G=2$, $\bar{P}_0=10$dB, $\sigma_{\sf ant}^2=\sigma_{\sf user}^2=0.25$.}
\label{fig:4}
\end{figure}

In this simulation,  we show how the worst user's SINR scales with the number of users served in the relay system. We consider the problem formulation with one total power constraint present. In Figure \ref{fig:4}, we have $L=8$, $G=2$, $\sigma_{\sf ant}^2=\sigma_{\sf user}^2=0.25$ and the total power threshold at relays is set to be $\bar P_0=10$dB.
From the figure, we see that (R2SDR) serves as an upper bound of (R1SDR), which is consistent with Propositions 2 and 3. Moreover, it shows that BFA AF can own better SINR performance than the BF AF scheme for all the values of $M$, which verifies the results in Proposition 1 and Theorem \ref{thm:main}. 

%\begin{figure}[htb]
%\centering
%\includegraphics[width=0.38\textwidth]{./r2relaytsp/MIMO_MM_MMF_82M_0db_10db_025_025_100.eps}
%\caption{Worst user's SINR versus number of users in the MIMO relay system: $L=8$, $G=2$, $\bar{P}_0=10$dB, $\sigma_{\sf ant}^2=\sigma_{\sf user}^2=0.25$.}
%\label{fig:3}
%\end{figure}

\vspace{-\baselineskip}

\subsection{Worst User's SINR versus Number of Per-relay Power Constraints}
\begin{figure}[htb]
\centering
\includegraphics[width=0.42\textwidth]{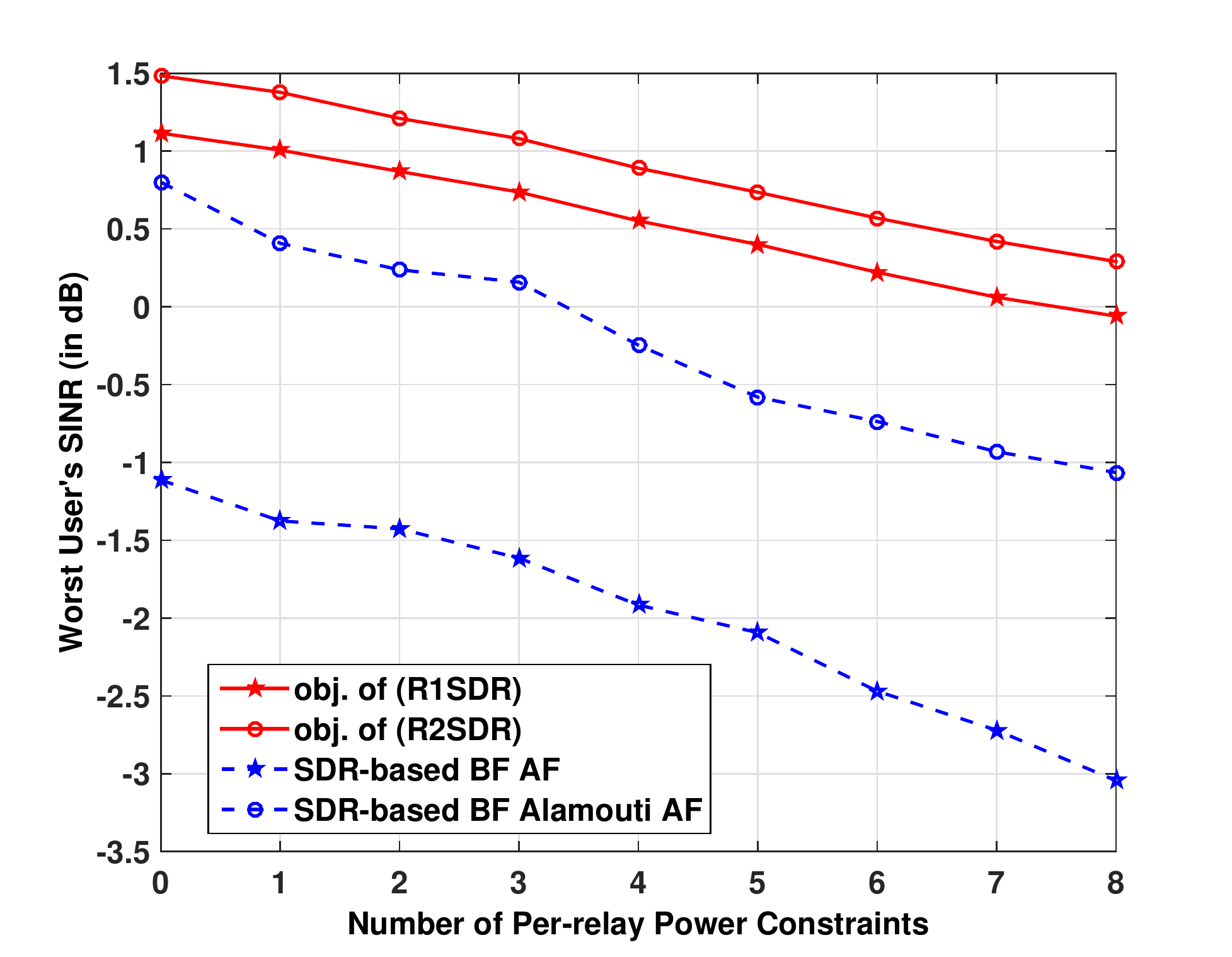}
\caption{Worst user's rate achieved by different AF schemes versus number of per-antenna power constraints in the distributed relay network: $L=8$, $G=2$, $M=16$, $\bar{P}_0=7$dB, $\bar{P}_\ell = -5$dB for $\ell=1,\ldots,L$, $\sigma_{\sf ant}^2=\sigma_{\sf user}^2=0.25$.}
\label{fig:6}
\end{figure}

In this simulation, we consider the scenario where both the total power constraint and per-antenna power constraints are present and the primal users are absent.  Our purpose is to see how the worst user's SINR scales with the number of per-relay power constraints. Specifically, Figure \ref{fig:6} shows the distributed relay case with $L=8$, $G=2$, $M=16$, where the total power threshold is $\bar{P}_0=7$dB and per-relay power threshold is also $\bar{P}_\ell=-5$dB. We set $\sigma_{\sf ant}^2=\sigma_{\sf user}^2=0.25$ and vary the number of per-relay power constraints from $0$ to $L$ to compare SINR performances of different AF schemes.  It shows that the BF Alamouti AF scheme outperforms the BF AF scheme.  As the number of per-relay power constraints increases, the SINRs diverge from their SDR upper bounds, and both BF AF and BF Alamouti AF exhibit the same scaling with $L$, which is consistent with the approximation bounds in terms of $J$ in Proposition 1 and Theorem 1.

\subsection{Worst User's SINR versus Number of Primal Users}
\begin{figure}[htb]
\centering
\includegraphics[width=0.42\textwidth]{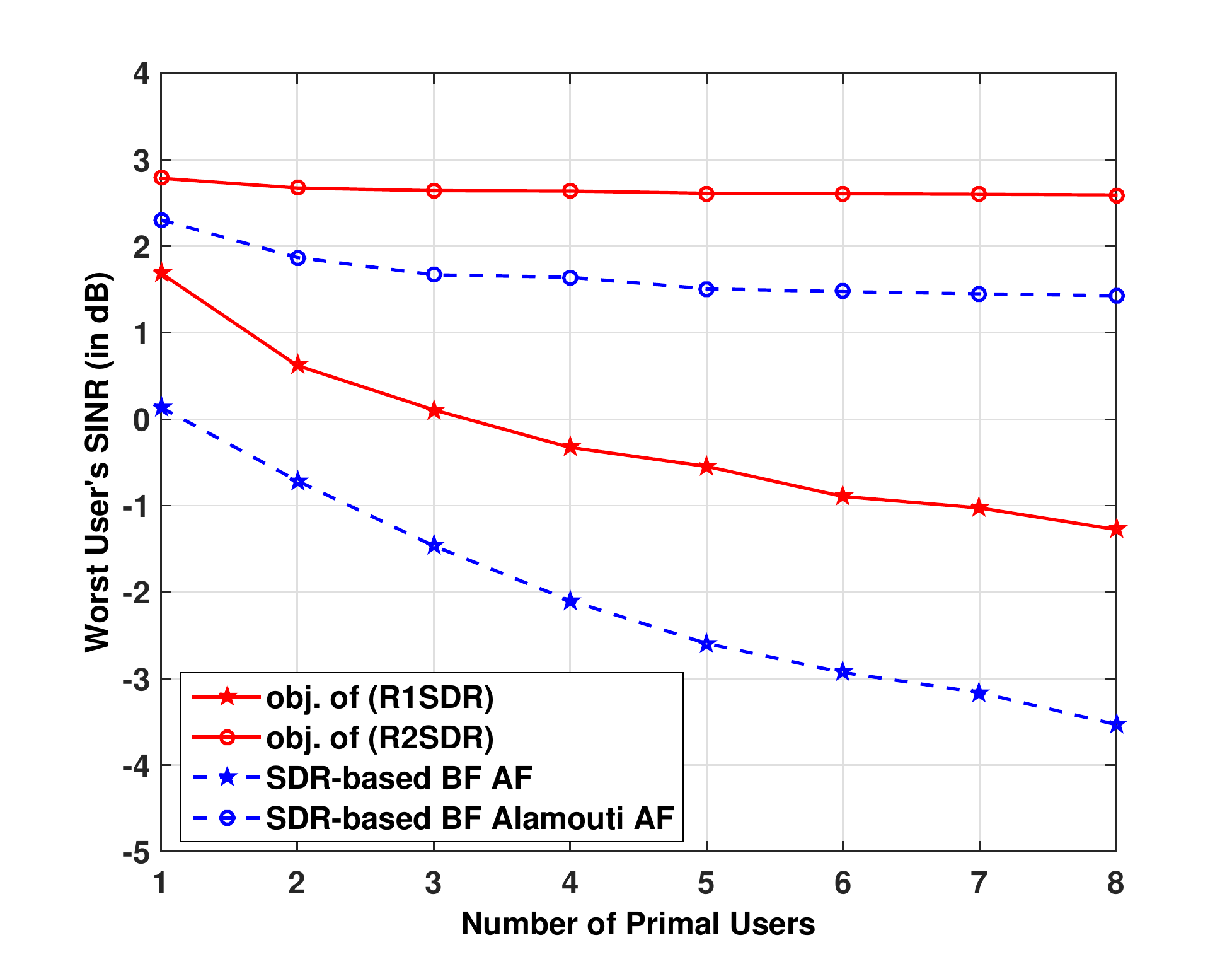}
\caption{Worst user's SINR versus number of primal users in the distributed relay CR network: $L=8$, $G=2$, $M=12$, $\bar{P}_0=10$dB, $b_u = 3$dB for $u=1,\ldots,U$, $\sigma_{\sf ant}^2=\sigma_{\sf user}^2=0.25$ and $\sigma_{\sf u}^2=0.25$.}
\label{fig:8}
\end{figure}

Similar to previous simulations, here we show the worst user's SINR scaling with the number of primal users. To set up the problem, we consider the scenario where the total power constraint and the primal users' interference constraints are present.  We assume that $L=8$, $G=2$ and $M=12$ and set $\sigma_{\sf ant}^2=\sigma_{\sf user}^2=0.25$,  the total power budget $\bar{P}_0=10$dB, and the noise power at all primal users to be $\sigma_{\sf u}^2=0.25$. Moreover, we assume that the primal users are subject to an interference power threshold equaling to $b_u=3$dB. We then increase the number of primal users to see the SINR scaling in Figure~\ref{fig:8}. It shows that as the primal users increases, both the BF AF scheme and BFA AF scheme diverge from their SDR bounds and BFA AF shows a significant improvement over BF AF. These results further validate Proposition 1 and Theorem 1 in terms of the scaling of $J$.

%\begin{figure}[htb]
%\centering
%\includegraphics[width=0.38\textwidth]{./r2relaytsp/cr_MIMO_MM_MMF_426_0db_10db_3db_025_025_025_100_increasePU.eps}
%\caption{Worst user's SINR versus number of primal users in the MIMO relay CR network: $L=4$, $G=2$, $M=12$, $\bar{P}_0=10$dB, $b_u = 3$dB for $u=1,\ldots,U$, $\sigma_{\sf ant}^2=\sigma_{\sf user}^2=0.25$ and $\sigma_{\sf u}^2=0.25$.}
%\label{fig:7}
%\end{figure}

\subsection{Simulation Results for Multicasting Relay Networks} \label{subsec:DRN}
\begin{figure}[htb]
\centering
\includegraphics[width=0.41\textwidth]{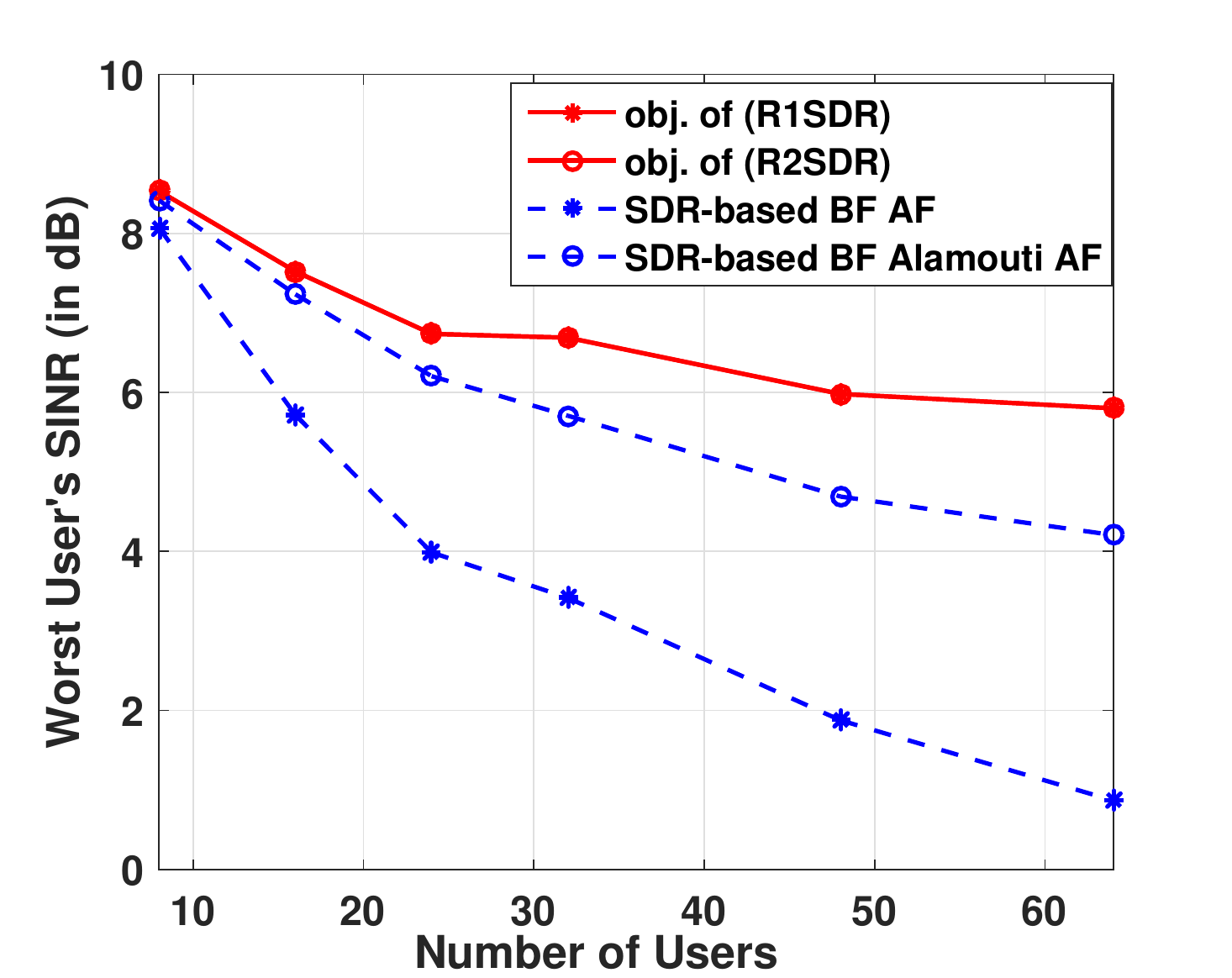}
\caption{Worst user's SINR versus the number of users in the distributed relay multicast network: $L=8$, $G=1$, $M=16$, $\bar{P}_0=10$dB, $b_u = 3$dB, $\sigma_{\sf ant}^2=\sigma_{\sf user}^2=0.25$ and $\sigma_{\sf u}^2=0.25$.}
\label{fig:11}
\end{figure}

\begin{figure}[htb]
\centering
\includegraphics[width=0.41\textwidth]{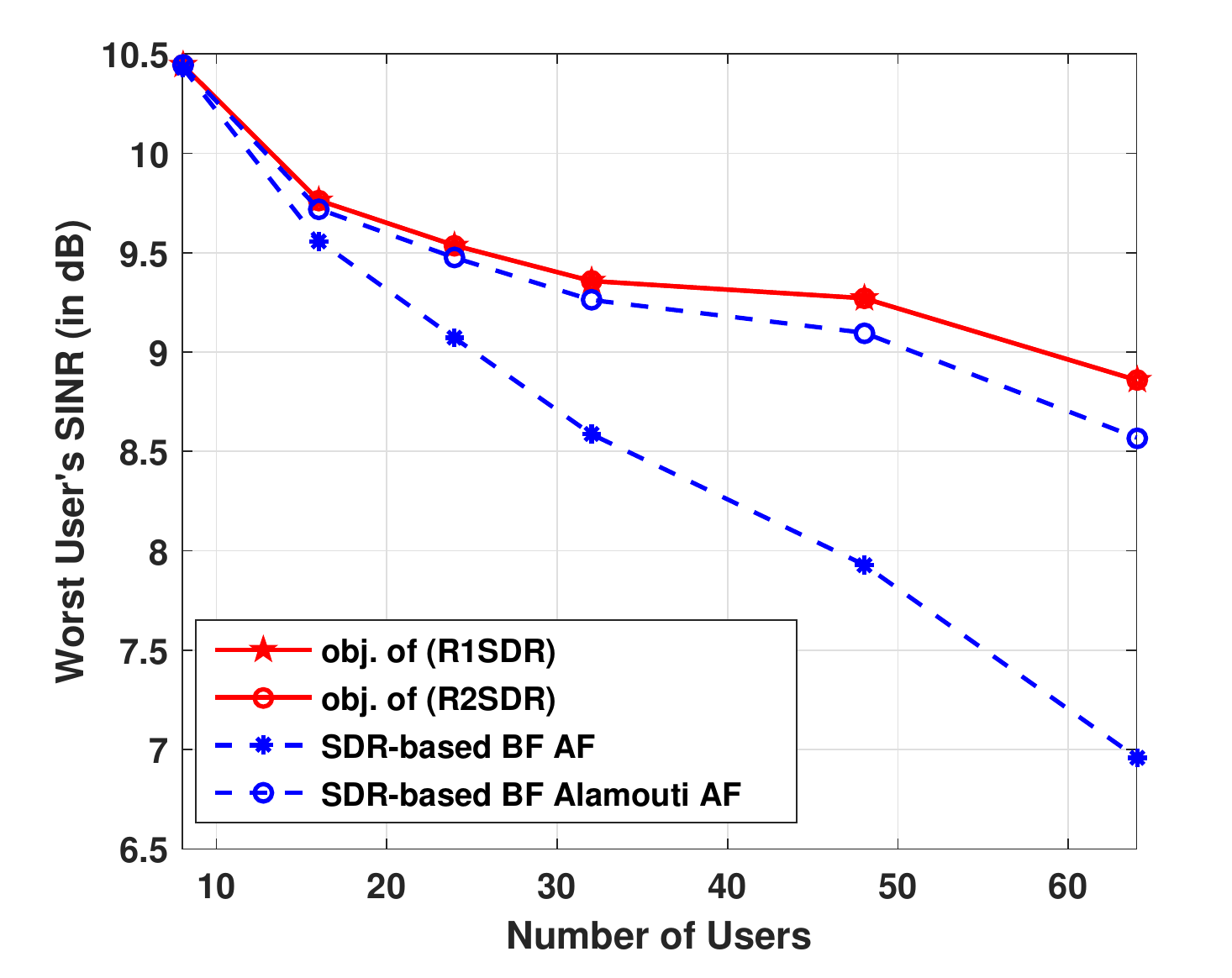}
\caption{Worst user's SINR versus the number of users in the MIMO relay multicast network: $L=4$, $G=1$, $M=16$, $\bar{P}_0=10$dB, $b_u = 3$dB, $\sigma_{\sf ant}^2=\sigma_{\sf user}^2=0.25$ and $\sigma_{\sf u}^2=0.25$}
\label{fig:12}
\end{figure}

In this sub-section, we provide numerical results for the special cases of multicasting. In both Figures \ref{fig:11} and \ref{fig:12}, we consider the problem formulation with the total power constraint and one primal user's temperature constraint and show the worst user's SINR scaling with the number of users. Specifically, in Figure \ref{fig:11}, we set $L=8$, $\bar P_0 = 10$dB and $\sigma_{\sf ant}^2=\sigma_{\sf user}^2=0.25$. The simulation results show that the obj. of (R1SDR) and (R2SDR) coincide with each other, and the curve of BFA AF scales better than that for BF AF, which is consistent with results in Proposition \ref{Prop:mimomulticast0}.  In Figure \ref{fig:12}, we have $L=4$, $\bar P_0 = 10$dB and $\sigma_{\sf ant}^2=\sigma_{\sf user}^2=0.25$. It shows that the obj. of (R2SDR) equal to that of (R1SDR). Moreover, BFA AF exhibits better scaling than BF AF as the number of users increases. This verifies results in Proposition \ref{Thm2}.

\subsection{Actual Bit Error Rate (BER) Performance} \label{subsec:BER}
\begin{figure}[htb]
\centering
\includegraphics[width=0.42\textwidth]{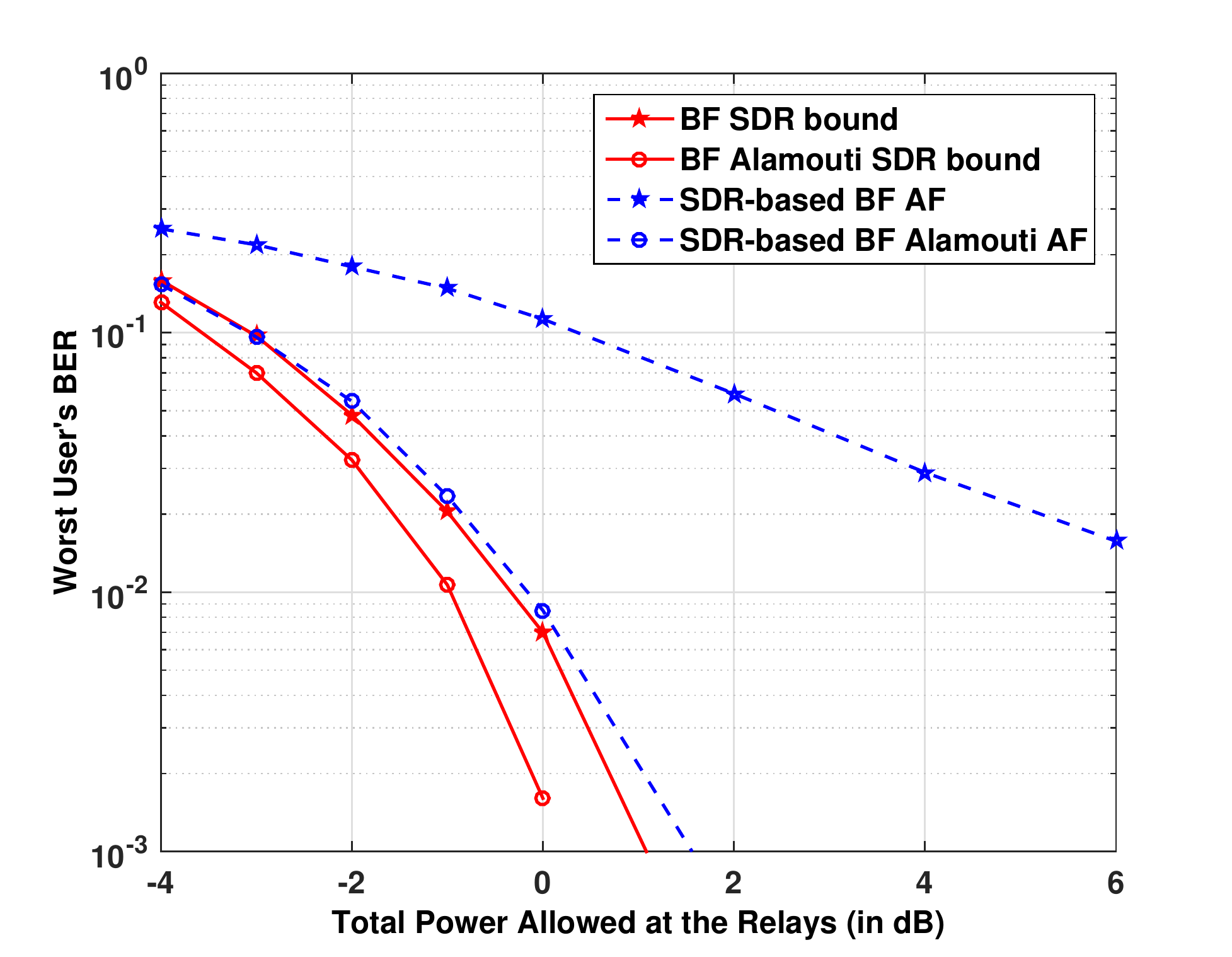}
\caption{Worst user's BER achieved by different AF schemes versus total power threshold in the distributed relay network: $L=8$, $G=2$, $M=12$, $\sigma_{\sf ant}^2=\sigma_{\sf user}^2=0.25$. A rate-$\frac{1}{3}$ turbo code with codelength $2880$ is used.}
\label{fig:14}
\end{figure}

To further demonstrate the efficacy of the proposed AF scheme, we study the actual coded bit error rate (BER) performance of the scenario setting in Figure \ref{fig:2}.  Specifically, for each time slot, we simulate the actual AF process by generating $s_j(t), {n}^\ell(t)$ according to the SISO model in~\eqref{yt} and \eqref{chapter2:eq:model_alam_uni_relay_mimo} and detecting and decoding $s_j(t)$ at each receivers.  The resulting BERs are shown in Figures~\ref{fig:14}.  To simulate the SDR bound in the BER plots, we assume that there exists an SISO channel whose SINR is equal to $\gamma({\bm W}^\star)$ or $\theta({\bm W}_1^\star, {\bm W}_2^\star)$. In our simulations, we adopt a gray-coded QPSK modulation scheme and a rate-$1/3$ turbo code in \cite{Std:16e} with a codelength of $2880$ bits.  We simulate $100$ code blocks for each channel realization and thus the BER reliability level is $10$e$-4$. We see that the actual BER performance of the proposed BFA AF scheme indeed outperforms the BF-AF scheme at almost all power thresholds.  The results are consistent with those SINR results in Figure~\ref{fig:2} and show that BFA AF can achieve a good performance in real applications.

%\begin{figure}[htb]
%\centering
%\includegraphics[width=0.38\textwidth]{./r2relaytsp/BER_r2_MIMO_MM_MMF_428_0db_025_025_100_2880.eps}
%\caption{Worst user's BER achieved by different AF schemes versus total power threshold at the MIMO relay: $L=4$, $G=2$, $M=16$, $\sigma_{\sf ant}^2=\sigma_{\sf user}^2=0.25$. A rate-$\frac{1}{3}$ turbo code with codelength $2880$ is used.}
%\label{fig:13}
%\end{figure}

\vspace{-0.5cm}
\section{Conclusions} \label{sec:conclusions}

In this work, we have studied the AF design problem in both the MIMO relay network and the distributed relay network. We considered different power constraints and tried to maximize the worst user's SINR under these power constraints in the multigroup multicasting scenario. A novel BFA AF scheme was proposed, for which the design problem is formulated as a two-variable fractional QCQP. Our main contributions lie in two aspects. First, we analyzed the new two-variable fractional SDR problem. The results show that by introducing $2$-DoF in the AF structure, we can increase the optimal objective value of the SDR problem. Moreover, the quality the SDR solution also improves, and therefore system performance can be enhanced.
Second, we showed in some special cases of relay multicasting that how the introduced one more DoF benefits the system performance and derived some interesting results, which can facilitate the system design. In summary, we demonstrate from both theoretical and numerically ways that the proposed BFA AF scheme can outperform the traditional BF AF scheme, especially in large-scale systems. Our future work is the AF design under imperfect CSI scenarios and other types of relay cloud networks with limited link capacity.
\appendix

\subsection{Proof of Theorem 1}\label{Thm1}
In this appendix, we provide the proof of Theorem 1 in four steps. Specifically, we begin by considering a specific randomization time $n$ in Algorithm \ref{alg:1}. W.l.o.g, we let ${\bm R}_j$ and $\bar{\bm R}_j$ denote the matrices associated with the power constraints, e.g., ${\bm R}_j={\bm G}_u, \bar{\bm R}_j=\bar{\bm G}_u$ or ${\bm R}_j=\bar{\bm R}_j={\bm D}_\ell$. Hence, herein $J$ is the total number of generalized power constraints.
Letting ${\bm W}_1 = (\bm{\xi}_1^n)({\bm{\xi}_1^n})^H$ and ${\bm W}_2 = (\bm{\xi}_2^n)({\bm{\xi}_2^n})^H$, 
our purpose in Steps 1-3 is to prove that for any proper $\rho, v>0$, the probability
\begin{align*}
&\mathcal{P} \triangleq \Pr \Big(
\left\{
\theta\left( {\bm W}_{1}, {\bm W}_{2} \right) \ge \rho \cdot \theta\left( {\bm W}_{1}^\star, {\bm W}_{2}^\star \right) \right\} \\\notag
&\wedge
\Big\{{\bm R}_j \bullet {\bm W}_1 + \bar{\bm R}_j \bullet {\bm W}_2 \le v ({\bm R}_j \bullet {\bm W}_1^\star + \bar {\bm R}_j \bullet {\bm W}_2^\star ), \\\notag
&\quad \forall j=1,...,J \Big\}\Big)
\end{align*}
is always positive.  Based on this result, in Step 4, we could further identify the approximation bounds.

\emph{{Step 1:} } To proceed the proof for $\mathcal{P}>0$, we define
\[
\theta_{k, i} ({\bm X}_1, {\bm X}_2) = \frac{{\bm A}_{k,i} \bullet {\bm X}_1 + \bar{\bm A}_{k,i} \bullet {\bm X}_2}{{\bm C}_{k,i} \bullet {\bm X}_1 + \bar{\bm C}_{k,i} \bullet {\bm X}_2 + 1}.
\]
Then, let us consider the events
\begin{align*}
&\mathcal{E}_{k,i} = \left\{ \theta_{k, i} ({\bm W}_1, {\bm W}_2) \le \rho \cdot \theta_{k, i} ({\bm W}_1^\star, {\bm W}_2^\star )\right\}, \\\notag
&\mathcal{F}_{j} =\left\{ {\bm R}_j \bullet {\bm W}_1 + \bar {\bm R}_j \bullet {\bm W}_2 \ge v \left({\bm R}_j \bullet {\bm W}_1^\star + \bar {\bm R}_j \bullet {\bm W}_2^\star \right) \right\},
\end{align*}
and define $\mathcal{E}_{k,i}^c$ and $\mathcal{F}_{j}^c$ to be the complement of $\mathcal{E}_{k,i}$ and $\mathcal{F}_{j}$. It is easy to see that
\begin{align*}
\mathcal{P}
%\ge &\Pr\left( \left( \bigcap_{k=1,\ldots,G \atop i=1,\ldots,m_k} \mathcal{E}_{k,i}^c \right) \cap \left( \bigcap_{j=1}^J \mathcal{F}_\ell^c \right) \right) \\
\ge &1 - \Pr\left( \bigcup_{k=1,\ldots,G \atop i=1,\ldots,m_k} \mathcal{E}_{k,i} \right) - \Pr\left( \bigcup_{j=1}^{J} \mathcal{F}_j \right).
\end{align*}

\emph{Step 2:} We intend to use the following Lemma \ref{lem:3} to bound $\Pr\left( \bigcup_{k=1,\ldots,G \atop i=1,\ldots,m_k} \mathcal{E}_{k,i} \right)$.
\begin{Lemma}
\label{lem:3} 
Given Hermitian positive semidefinite matrices ${\bm A}$, $\bar{\bm A}$,  ${\bm C}$, $\bar{\bm C}$ with ${\rm rank}({\bm A}) ={\rm rank}(\bar{\bm A})=1$, let ${\bm \xi} \sim \mathcal{CN}({\bf 0}, {\bm X}_1^\star)$, ${\bm \eta} \sim \mathcal{CN}({\bf 0}, {\bf X}_2^\star)$ be independent random vectors. Assume that ${\bm A}\bullet{\bm X}_1^\star>0$ and $\bar{\bm A}\bullet{\bm X}_2^\star>0$. 
Then, 

\noindent
\emph{1).} For any $\rho <1/2$, it holds that
\begin{align}\notag
&\Pr\left( 
\frac{{\bm \xi}^H{\bm A}{\bm \xi} + {\bm \eta}^H\bar{\bm A}{\bm \eta}} {{\bm \xi}^H{\bm C}{\bm \xi} +  {\bm \eta}^H\bar{\bm C} {\bm \eta} + 1}\le \rho \frac{{\bm A} \bullet {\bm X}_1^\star + \bar{\bm A} \bullet {\bm X}_2^\star}{{\bm C}\bullet {\bm X}_1^\star + \bar{\bm C} \bullet {\bm X}_2^\star + 1}
\right) \\\label{twoupbds1}
&\le \frac{4\rho}{1-2\rho},
\end{align}

\noindent
\emph{2).} For for any $\rho < \omega/2$ where $\omega = \frac{\min\{{\bm A}\bullet{\bm X}_1^\star,~\bar{\bm A}\bullet{\bm X}_2^\star\}}{{\bm A}\bullet{\bm X}_1^\star + \bar{\bm A}\bullet{\bm X}_2^\star}$, it holds that
\begin{align}\notag
&\Pr\left( 
\frac{{\bm \xi}^H{\bm A}{\bm \xi} + {\bm \eta}^H\bar{\bm A}{\bm \eta}} {{\bm \xi}^H{\bm C}{\bm \xi} +  {\bm \eta}^H\bar{\bm C} {\bm \eta} + 1}\le \rho \frac{{\bm A} \bullet {\bm X}_1^\star + \bar{\bm A} \bullet {\bm X}_2^\star}{{\bm C}\bullet {\bm X}_1^\star + \bar{\bm C} \bullet {\bm X}_2^\star + 1}
\right) \\\label{twoupbds2}
&\le \left(\frac{4\rho}{\omega-2\rho}\right)^2.
\end{align}
\end{Lemma}
\emph{Proof:} We delegate the proof in the 
the companion technical report~\cite{CompTechRepr2}. We remark that the upper bounds in \eqref{twoupbds1} and \eqref{twoupbds2} hold simultaneously. 

In our problem, we have ${\rm rank}({\bm A}_{k,i})={\rm rank}(\bar{\bm A}_{k,i})=1$. Denoting  $c_{k,i} = \frac{\min\{{\bm A}_{k, i}\bullet{\bm W}_1^\star, ~\bar{\bm A}_{k, i}\bullet{\bm W}_2^\star\}}{{\bm A}_{k, i}\bullet{\bm W}_1^\star + \bar{\bm A}_{k, i}\bullet{\bm W}_2^\star}$, we can apply \eqref{twoupbds1} to bound event $\mathcal{E}_{k, i}$ as
$
\Pr\left(\mathcal{E}_{k, i}\right)\le  \left( \frac{4\rho}{c_{k,i}-2\rho} \right)^2, ~\forall k, i.
$
Then, by choosing $\rho=\omega/(7\sqrt{M})$ with
$
\omega = \displaystyle\min_{k, i}\left\{c_{k, i}\right\}
$,
we may apply Lemma \ref{lem:3} to obtain
\begin{align}\notag
\Pr\left(\bigcup_{k=1,\ldots,G \atop i=1,\ldots,m_k} \mathcal{E}_{k,i}\right) &\le \sum_{k, i}\Pr\left(\mathcal{E}_{k, i}\right)\le  M\left( \frac{4\rho}{c_{k,i}-2\rho} \right)^2 < \frac{3}{4}.
\end{align}
On the other hand, we can also consider the second upper bound in Lemma \ref{lem:3}.
By choosing $\rho=1/(8M)$ we have
\begin{align}\notag
\Pr\left(\bigcup_{k=1,\ldots,G \atop i=1,\ldots,m_k} \mathcal{E}_{k,i}\right) &\le \sum_{k, i}\Pr\left(\mathcal{E}_{k, i}\right)\le  M \frac{4\rho}{1-2\rho} < \frac{3}{4}.
\end{align}
Note that these two bounds hold simultaneously given that we pick different values for $\rho$.

\emph{Step 3:} We then bound $\Pr\left( \bigcup_{j=1}^{J} \mathcal{F}_j \right)$ by Lemma \ref{lem:4}.
\begin{Lemma}
\label{lem:4} 
Let ${\bm D}$ and $\bar{\bm D}$ be Hermitian positive semidefinite matrices and ${\bm \xi} \sim \mathcal{CN}({\bm 0}, {\bm X}_1^\star)$, ${\bm \eta} \sim \mathcal{CN}({\bf 0}, {\bm X}_2^\star)$ be independent random vectors. If $\bm{D} \bullet \bm{X}_1^\star + \bar{\bm{D}} \bullet \bm{X}_2^\star=0$, then ${\bm \xi}^H\bm{D}{\bm \xi} + {\bm \eta}^H\bar{\bm{D}}{\bm \eta}= 0$ almost surely. Otherwise, for any $v \ge 2$, we have
\begin{align*}
&\Pr\left( {\bm \xi}^H{\bm D}{\bm \xi} + {\bm \eta}^H\bar{\bm D}{\bm \eta}  \ge v ({\bm D} \bullet {\bm X}_1^\star + \bar{\bm D} \bullet {\bm X}_2^\star  ) \right)\le 2\exp(-\frac{v}{2}).
\end{align*}
\end{Lemma}
\emph{Proof:} We delegate the proof in in the 
the companion technical report~\cite{CompTechRepr2}. 

\noindent
Similar to Step 2,  by choosing $v= 2\log (16J)$, we can bound $\Pr\left( \bigcup_{j=1}^{J} \mathcal{F}_j \right)$ as
\begin{align}\notag
\Pr\left( \bigcup_{j=1}^{J} \mathcal{F}_j \right) &\le \sum_{j=1}^{J} \Pr\left(\mathcal{F}_j \right) \le 2J\cdot \exp(-\frac{v}{2}) =\frac{1}{8}.
\end{align}
In this way, we can bound the probability $\mathcal{P}$ as
\begin{align*}
\mathcal{P} \ge 1 - \Pr\left( \bigcup_{k=1,\ldots,G \atop i=1,\ldots,m_k} \mathcal{E}_{k,i} \right) - \Pr\left( \bigcup_{\ell=0}^{L} \mathcal{F}_\ell \right)>\frac{1}{8}.
\end{align*}

\emph{Step 4:} Now it is ready for us to show that based on the fact $\mathcal{P}>0$, the desired approximation bounds can be derived.  
In particular, letting $\widehat{\bm{W}}_1 ={\bm{W}}_1/v$ and $\widehat{\bm{W}}_2={\bm{W}}_2/v$, previous discussions imply that with probability at least $1/8$, the rank-one solutions $\left(\widehat{\bm{W}}_1, \widehat{\bm{W}}_2\right)$ are feasible for Problem (R2SDR) and we have
\begin{align}\notag
\theta\left( \widehat{\bm{W}}_1, \widehat{\bm{W}}_2  \right)=&\min_{k=1,\ldots,G \atop i=1,\ldots,m_k }\frac{{\bm A}_{k,i}\bullet ({\bm W}_1/v) + \bar{\bm A}_{k,i}\bullet ({\bm W}_2/v)} {{\bm C}_{k,i}\bullet ({\bm W}_1/v)+\bar{\bm C}_{k,i}\bullet ({\bm W}_2/v)+1} \\\notag
%=&\min_{k=1,\ldots,G \atop i=1,\ldots,m_k }\frac{{\bm A}_{k,i}\bullet{\bm W}_1 + \bar{\bm A}_{k,i}\bullet{\bm W}_2} {{\bm C}_{k,i}\bullet {\bm W}+\bar{\bm C}_{k,i}\bullet{\bm W}_2+1} \\\notag
%&\cdot \frac{{\bm C}_{k,i}\bullet {\bm W}_1+\bar{\bm C}_{k,i}\bullet{\bm W}_2+1} {{\bm C}_{k,i}\bullet {\bm W}_1+\bar{\bm C}_{k,i}\bullet{\bm W}_2+v} \\\notag
%\ge&\frac{1}{v} \min_{k=1,\ldots,G \atop i=1,\ldots,m_k }\frac{{\bm A}_{k,i}\bullet{\bm W}_1 + \bar{\bm A}_{k,i}\bullet{\bm W}_2} {{\bm C}_{k,i}\bullet {\bm W}_1+\bar{\bm C}_{k,i}\bullet{\bm W}_2+1} \\\notag
\ge&\frac{\rho}{v}\cdot \theta\left( {\bm{W}_1^\star}, {\bm{W}_2^\star} \right).
\end{align}
Therefore, we have
\begin{equation}\label{r2}
\theta\left( \widehat{\bm{W}}_1, \widehat{\bm{W}}_2 \right)  \ge \frac{\omega}{7\sqrt{M}(2\log (16J)+1)} \cdot \theta\left({\bm{W}}_1^\star, {\bm{W}}_2^\star\right)
\end{equation}
and
\begin{equation}\label{r1}
\theta\left( \widehat{\bm{W}}_1, \widehat{\bm{W}}_2 \right)  \ge \frac{1}{8M(2\log (16J)+1)}\cdot \theta\left({\bm{W}}_1^\star, {\bm{W}}_2^\star\right)
\end{equation}
hold simultaneously.
Since \eqref{r2} and \eqref{r1} are true for each randomization $n\in\{1,\ldots,N\}$, it follows that 
\begin{align*}\label{thm1}
& \Pr \left( \left\{ \exists n: \theta\left( \widehat{\bm{w}}_1^n(\widehat{\bm{w}}_1^n)^H,  \widehat{\bm{w}}_2^n(\widehat{\bm{w}}_2^n)^H \right)
\ge c \cdot \theta\left( {\bm{W}_1^\star},  {\bm{W}_2^\star}\right) \right\} \right)\\
&\ge 1-(7/8)^N,
\end{align*}
where $\widehat{\bm{w}}_1^n$ and $\widehat{\bm{w}}_2^n$ are extracted by eigen-decomposition from rank-one matrices $\widehat{\bm{W}}_1$ and $\widehat{\bm{W}}_2$ at randomization time $n$, and $c$ is defined in Theorem 1.
This,  together with \eqref{eq:W_w}, completes the proof of Theorem 1. \hfill $\blacksquare$

% Generated by IEEEtran.bst, version: 1.14 (2015/08/26)

%\bibliographystyle{IEEEbib}
%\bibliography{IEEEabrv,robust_r2_relay_ref}

\begin{thebibliography}{10}
\providecommand{\url}[1]{#1}
\csname url@samestyle\endcsname
\providecommand{\newblock}{\relax}
\providecommand{\bibinfo}[2]{#2}
\providecommand{\BIBentrySTDinterwordspacing}{\spaceskip=0pt\relax}
\providecommand{\BIBentryALTinterwordstretchfactor}{4}
\providecommand{\BIBentryALTinterwordspacing}{\spaceskip=\fontdimen2\font plus
\BIBentryALTinterwordstretchfactor\fontdimen3\font minus
  \fontdimen4\font\relax}
\providecommand{\BIBforeignlanguage}[2]{{%
\expandafter\ifx\csname l@#1\endcsname\relax
\typeout{** WARNING: IEEEtran.bst: No hyphenation pattern has been}%
\typeout{** loaded for the language `#1'. Using the pattern for}%
\typeout{** the default language instead.}%
\else
\language=\csname l@#1\endcsname
\fi
#2}}
\providecommand{\BIBdecl}{\relax}
\BIBdecl

\bibitem{ITAWIFI}
J.~Chamberland, P.~Eedara, and A.~Taghavi, ``Occupancy estimation using
  distributed {WiFi} monitoring,'' in \emph{ITA Workshop 2016}, Feb. 2016.

\bibitem{tehrani2014device}
M.~N. Tehrani, M.~Uysal, and H.~Yanikomeroglu, ``Device-to-device communication
  in {5G} cellular networks: challenges, solutions, and future directions,''
  \emph{IEEE Communications Magazine}, vol.~52, no.~5, pp. 86--92, 2014.

\bibitem{cloudrelay0}
T.~Hynek, D.~Halls, and J.~Sykora, ``Practical implementation of cloud
  initialization procedure for wireless physical layer network coding clouds,''
  in \emph{20th European Wireless Conference; Proceedings of European Wireless
  2014}, May 2014, pp. 1--6.

\bibitem{li2014social}
Y.~Li, T.~Wu, P.~Hui, D.~Jin, and S.~Chen, ``Social-aware {D2D} communications:
  qualitative insights and quantitative analysis,'' \emph{Communications
  Magazine, IEEE}, vol.~52, no.~6, pp. 150--158, 2014.

\bibitem{wpcran}
K.~C. et~al., ``{C-RAN}: the road toward green {RAN}, white paper,'' in
  \emph{China Mobile Research Institute}, 2011.

\bibitem{shi2013group}
Y.~Shi, J.~Zhang, and K.~Letaief, ``Group sparse beamforming for green
  {Cloud-RAN},'' \emph{{IEEE} Trans. Wireless Commun.}, May 2014.

\bibitem{cranfronthaul}
V.~N. Ha, L.~B. Le, and N.-D. Dao, ``Cooperative transmission in cloud {RAN}
  considering fronthaul capacity and cloud processing constraints,'' in
  \emph{Proc. IEEE Wireless Communications and Networking Conference (WCNC)},
  2014.

\bibitem{wucrn_r2}
S.~X. Wu, A.~M.-C. So, and W.-K. Ma, ``A beamformed {Alamouti}
  amplify-and-forward scheme in multigroup multicast cloud-relay networks,'' in
  \emph{Proc. IEEE Int. Conf. Acoust., Speech, Signal Process. (ICASSP)}, 2015.

\bibitem{WuLiMaSo2015SPAWC}
S.~X. Wu, Q.~Li, W.-K. Ma, and A.~M.-C. So, ``Stochastic amplify-and-forward
  schemes for multigroup multicast transmission in a distributed relay
  network,'' in \emph{Proc.~15th IEEE Int. Workshop Signal Process. Advances in
  Wireless Commun. (SPAWC)}, Jun. 2015.

\bibitem{wu2015relaysbf}
S.~X. Wu, Q.~Li, A.~M.-C.So, and W.-K. Ma, ``A stochastic beamformed
  amplify-and-forward scheme in a multigroup multicast {MIMO} relay network
  with per-antenna power constraints,'' \emph{submitted to {IEEE} Trans.
  Wireless Commun.}, 2015, available online at {\urlstyle{tt}
  \url{http://arxiv.org/abs/1510.02448}}.

\bibitem{fazeli2009multiple}
S.~Fazeli-Dehkordy, S.~Shahbazpanahi, and S.~Gazor, ``Multiple peer-to-peer
  communications using a network of relays,'' \emph{IEEE Trans. Signal
  Process.}, vol.~57, no.~8, pp. 3053--3062, 2009.

\bibitem{Jnl:Relay_Chalise_09}
B.~K. Chalise and L.~Vandendorpe, ``{MIMO} relay design for
  multipoint-to-multipoint communications with imperfect channel state
  information,'' \emph{{IEEE} Trans. Signal Process.}, vol.~57, no.~7, pp.
  2785--2796, 2009.

\bibitem{chalise2007mimo}
B.~K. Chalise, L.~Vandendorpe, and J.~Louveaux, ``{MIMO} relaying for
  multi-point to multi-point communication in wireless networks,'' in
  \emph{Computational Advances in Multi-Sensor Adaptive Processing, 2007.
  CAMPSAP 2007. 2nd IEEE International Workshop on}.\hskip 1em plus 0.5em minus
  0.4em\relax IEEE, 2007, pp. 217--220.

\bibitem{distributed_relay_Ding_08}
Z.~Ding, W.~H. Chin, and K.~K. Leung, ``Distributed beamforming and power
  allocation for cooperative networks,'' \emph{{IEEE} Trans. Wireless Commun.},
  vol.~7, no.~5, pp. 1817--1822, May 2008.

\bibitem{havary2008distributed}
V.~Havary-Nassab, S.~Shahbazpanahi, A.~Grami, and Z.-Q. Luo, ``Distributed
  beamforming for relay networks based on second-order statistics of the
  channel state information,'' \emph{{IEEE} Trans. Signal Process.}, vol.~56,
  no.~9, pp. 4306--4316, 2008.

\bibitem{jimenez2012non}
I.~Jimenez, M.~Barrenechea, M.~Mendicute, and E.~Arruti, ``Non-linear precoding
  approaches for non-regenerative multiuser {MIMO} relay systems,'' in
  \emph{Signal Processing Conference (EUSIPCO), 2012 Proceedings of the 20th
  European}.\hskip 1em plus 0.5em minus 0.4em\relax IEEE, 2012, pp. 1399--1403.

\bibitem{distributed_relay_Goma07}
K.~Gomadam and S.~Jafar, ``Optimal distributed beamforming in relay networks
  with common interference,'' in \emph{Global Telecommunications Conference,
  2007. GLOBECOM '07. IEEE}, Nov 2007, pp. 3868--3872.

\bibitem{khandaker2012joint}
M.~R. Khandaker and Y.~Rong, ``Joint transceiver optimization for multiuser
  {MIMO} relay communication systems,'' \emph{{IEEE} Trans. Signal Process.},
  vol.~60, no.~11, pp. 5977--5986, 2012.

\bibitem{KR14}
R.~Muhammad, A.~Khandaker, and Y.~Rong, ``Transceiver optimization for
  multi-hop {MIMO} relay multicasting from multiple sources,'' \emph{{IEEE}
  Trans. Wireless Commun.}, vol.~13, no.~9, pp. 5162--5172, Sep. 2014.

\bibitem{Jnl:MagzineMaLUO}
Z.-Q. Luo, W.-K. Ma, A.~M.-C. So, Y.~Ye, and S.~Zhang, ``Semidefinite
  relaxation of quadratic optimization problems,'' \emph{{IEEE} Signal Process.
  Mag.}, vol.~27, no.~3, pp. 20--34, May 2010.

\bibitem{chang2008approximation}
T.-H. Chang, Z.-Q. Luo, and C.-Y. Chi, ``Approximation bounds for semidefinite
  relaxation of max-min-fair multicast transmit beamforming problem,''
  \emph{IEEE Trans. Signal Process.}, vol.~56, no.~8, pp. 3932--3943, 2008.

\bibitem{jimulti13}
S.~Ji, S.~X. Wu, A.~M.~C. So, and W.-K. Ma, ``Multi-group multicast beamforming
  in cognitive radio networks via rank-two transmit beamformed {Alamouti}
  space-time coding,'' in \emph{Proc. IEEE Int. Conf. Acoust., Speech, and
  Signal Process. (ICASSP)}, May 2013, pp. 4409--4413.

\bibitem{MainPaper}
S.~X. Wu, W.-K. Ma, and A.~M.~C. So, ``Physical-layer multicasting by
  stochastic transmit beamforming and {Alamouti} space-time coding,''
  \emph{IEEE Trans. Signal Process.}, vol.~61, no.~17, pp. 4230--4245, Sept.
  2013.

\bibitem{schad2012convex}
A.~Schad, K.~L. Law, and M.~Pesavento, ``A convex inner approximation technique
  for rank-two beamforming in multicasting relay networks,'' in
  \emph{Proceedings of the 20th European Signal Processing Conference
  (EUSIPCO)}, 2012, pp. 1369--1373.

\bibitem{schad2015rank}
A.~Schad, K. L. Law and M. Pesavento, ``Rank-two beamforming and power allocation in multicasting relay
  networks,'' \emph{{IEEE} Trans. Signal Process.}, vol.~63, no.~13, pp.
  3435--3447, 2015.

\bibitem{bornhorst2012distributed}
N.~Bornhorst, M.~Pesavento, and A.~B. Gershman, ``Distributed beamforming for
  multi-group multicasting relay networks,'' \emph{{IEEE} Trans. Signal
  Process.}, vol.~60, no.~1, pp. 221--232, 2012.

\bibitem{tran2014conic}
L.-N. Tran, M.~F. Hanif, and M.~Juntti, ``A conic quadratic programming
  approach to physical layer multicasting for large-scale antenna arrays,''
  \emph{{IEEE} Signal Process. Lett.}, vol.~21, no.~1, pp. 114--117, 2014.

\bibitem{christopoulos2015multicast}
D.~Christopoulos, S.~Chatzinotas, and B.~Ottersten, ``Multicast multigroup
  beamforming for per-antenna power constrained large-scale arrays,''
  \emph{Proc. of IEEE Int. Work-shop on Signal Process. Advances for Wirel.
  Commun (SPAWC)}, Jun. 2015.

\bibitem{mehanna2015feasible}
O.~Mehanna, K.~Huang, B.~Gopalakrishnan, A.~Konar, and N.~Sidiropoulos,
  ``Feasible point pursuit and successive approximation of non-convex
  {QCQPs},'' \emph{{IEEE} Signal Process. Lett.}, vol.~22, no.~7, pp. 804--808,
  2015.

\bibitem{Gopalakrishnan15}
B.~Gopalakrishnan and N.~Sidiropoulos, ``High performance adaptive algorithms
  for single-group multicast beamforming,'' \emph{{IEEE} Trans. Signal
  Process.}, vol.~63, no.~16, pp. 4373--4384, Aug 2015.

\bibitem{CompTechRepr2}
S.~X. Wu, J.~Pan, A.~M.-C. So, and W.-K. Ma, ``Some proof derivations and
  further simulation results for {``Strategies for Exploring Two
  Degrees-of-freedom in Relay Beamforming Networks and the Performance
  Analysis''},'' \emph{Technical Report, Department of Systems Engineering and
  Engineering Management, The Chinese University of Hong Kong,} Feb 2016.
  Available online at {http://arxiv.org/abs/1602.06500}.

\bibitem{huang2012robust}
Y.~Huang, Q.~Li, W.-K. Ma, and S.~Zhang, ``Robust multicast beamforming for
  spectrum sharing-based cognitive radios,'' \emph{{IEEE} Trans. Signal
  Process.}, vol.~60, no.~1, pp. 527--533, 2012.

\bibitem{phan2009spectrum}
K.~T. Phan, S.~Vorobyov, N.~D. Sidiropoulos, C.~Tellambura \emph{et~al.},
  ``Spectrum sharing in wireless networks via qos-aware secondary multicast
  beamforming,'' \emph{IEEE Transactions on Signal Processing}, vol.~57, no.~6,
  pp. 2323--2335, 2009.

\bibitem{zhang2011optimal}
Y.~J. Zhang and A.~M.-C. So, ``Optimal spectrum sharing in mimo cognitive radio
  networks via semidefinite programming,'' \emph{IEEE Journal on Selected Areas
  in Communications}, vol.~29, no.~2, pp. 362--373, 2011.

\bibitem{Jnl:Karipidis_MM_2008}
E.~Karipidis, N.~D. Sidiropoulos, and Z.-Q. Luo, ``Quality of service and
  max-min-fair transmit beamforming to multiple co-channel multicast groups,''
  \emph{{IEEE} Trans. Signal Process.}, vol.~56, no.~3, pp. 1268--1279, Mar.
  2008.

\bibitem{MulticastLuo07}
Z.-Q. Luo, N.~D. Sidiropoulos, P.~Tseng, and S.~Zhang, ``Approximation bounds
  for quadratic optimization with homogeneous quadratic constraints,''
  \emph{SIAM J. Optim.}, vol.~18, no.~1, pp. 1--28, 2007.

\bibitem{Varshney}
L.~R. Varshney, ``Transporting information and energy simultaneously,'' in
  \emph{Information Theory, 2008. ISIT 2008. IEEE International Symposium
  on}.\hskip 1em plus 0.5em minus 0.4em\relax IEEE, 2008, pp. 1612--1616.

\bibitem{Grover}
P.~Grover and A.~Sahai, ``Shannon meets tesla: Wireless information and power
  transfer.'' in \emph{ISIT}, 2010, pp. 2363--2367.

\bibitem{Zhang}
R.~Zhang and C.~K. Ho, ``Mimo broadcasting for simultaneous wireless
  information and power transfer,'' \emph{Wireless Communications, IEEE
  Transactions on}, vol.~12, no.~5, pp. 1989--2001, 2013.

\bibitem{he2008semidefinite}
S.~He, Z.-Q. Luo, J.~Nie, and S.~Zhang, ``Semidefinite relaxation bounds for
  indefinite homogeneous quadratic optimization,'' \emph{SIAM Journal on
  Optimization}, vol.~19, no.~2, pp. 503--523, 2008.

\bibitem{tse2005fundamentals}
D.~Tse and P.~Viswanath, \emph{Fundamentals of wireless communication}.\hskip
  1em plus 0.5em minus 0.4em\relax Cambridge university press, 2005.

\bibitem{Jnl:Yongwei_rank}
Y.~Huang and S.~Zhang, ``Complex matrix decomposition and quadratic
  programming,'' \emph{Math. of Oper. Research}, vol.~32, pp. 758--768, 2007.

\bibitem{Std:16e}
\emph{{IEEE} Standard for Local and Metropolitan Area Networks, Part 16: Air
  Interface for Fixed and Mobile Broadband Wireless Access Systems}, IEEE Std.
  802.16e, 2005.

\end{thebibliography}
\end{document}